\documentclass[final,3p,times]{elsarticle}

\usepackage{graphicx}

\usepackage{amssymb}
\usepackage{amsmath}
\usepackage{subfig}
\usepackage{graphicx}
\usepackage{url}
\usepackage{algpseudocode}
\usepackage{algorithm}
\usepackage{multirow}
\usepackage{wasysym}
\usepackage{marvosym}
\usepackage{color, xcolor, colortbl}
\usepackage{array, booktabs}
\usepackage{xspace}
\usepackage{enumitem}

\usepackage{bm}

\newcommand{\secref}[1]{Section \ref{#1}}
\newcommand{\fref}[1]{Fig.~\ref{#1}}
\newcommand{\tref}[1]{Table~\ref{#1}}

\renewcommand{\eqref}[1]{Eq.~(\ref{#1})}
\newcommand{\eref}[1]{(\ref{#1})}

\newcommand{\mat}[1]{\textrm{\textbf{#1}}}

\newcommand{\psif}{\bm{\psi} (\bm{r}, \bm{\Omega})}
\newcommand{\psifd}{\bm{\psi} (\bm{r}, \bm{\Omega}')}

\newcommand{\xten}[1]{$\times$ 10$^{\text{#1}}$}

\definecolor{light}{rgb}{0.8,0.8,0.8}
\definecolor{medium}{rgb}{0.6,0.6,0.6}
\definecolor{dark}{rgb}{0.4,0.4,0.4}
\definecolor{darkmed}{rgb}{0.3,0.3,0.3}
\definecolor{darkest}{rgb}{0.2,0.2,0.2}
\definecolor{Black}{rgb}{0,0,0}
\definecolor{White}{rgb}{1,1,1}
\definecolor{lightpurple}{rgb}{0.78823,0.709803,0.74509}
\definecolor{lightpurpletext}{rgb}{0.788235,0.5529411,0.658823}
\definecolor{skyblue}{rgb}{0.80392,0.866666,0.92941}
\definecolor{skybluetext}{rgb}{0.61568627,0.7647058,0.913725}
\definecolor{darkgreen}{rgb}{0.3137254,0.458823,0.18431}
\definecolor{foliagegreen}{rgb}{0.188,0.415,0.105}
\definecolor{steelbluegrey}{rgb}{0.1961,0.2353,0.2392}
\definecolor{highlightblue}{rgb}{0.4078,0.6431,0.85}
\definecolor{matlabblue}{rgb}{0,0.2705,0.85}
\definecolor{darkred}{rgb}{0.8,0.1725,0}
\definecolor{fireenginered}{rgb}{0.505,0.1411,0}
\definecolor{darkpurple}{rgb}{0.6431,0.3137,0.8509}
\definecolor{gaylordpurple}{rgb}{0.416,0.204,0.549}
\definecolor{deludedorange}{rgb}{0.7409,0.4392,0}
\definecolor{darksalmon}{rgb}{0.9137,0.411,0.706}

\newcolumntype{a}{>{\columncolor{light}}c}
\newcolumntype{b}{>{\columncolor{skyblue}}c}

\makeatletter
\def\ps@pprintTitle{%
  \let\@oddhead\@empty
  \let\@evenhead\@empty
  \def\@oddfoot{\reset@font\hfil\thepage\hfil}
  \let\@evenfoot\@oddfoot
}
\makeatother

\begin{document}

\begin{frontmatter}



\title{Angular adaptivity in P$^0$ space and reduced tolerance solves for Boltzmann transport\tnoteref{crown}}
\author[AMCG]{S. Dargaville}
\ead{dargaville.steven@gmail.com}
\tnotetext[crown]{UK Ministry of Defence © Crown owned copyright 2023/AWE}
\author[AWE,AMCG]{R.P. Smedley-Stevenson}
\author[AMEC,AMCG]{P.N. Smith}
\author[AMCG]{C.C. Pain}
\address[AMCG]{Applied Modelling and Computation Group, Imperial College London, SW7 2AZ, UK}
\address[AWE]{AWE, Aldermaston, Reading, RG7 4PR, UK}
\address[AMEC]{ANSWERS Software Service, Jacobs, Kimmeridge House, Dorset Green Technology Park, Dorchester, DT2 8ZB, UK}
\begin{abstract}
Previously we developed an adaptive method in angle, based on solving in Haar wavelet space with a matrix-free multigrid for Boltzmann transport problems. This method scalably mapped to the underlying P$^0$ space during every matrix-free matrix-vector product, however the multigrid method itself was not scalable in the streaming limit. 

To tackle this we recently built an iterative method based on using an ideal restriction multigrid with frozen GMRES polynomials (AIRG) for Boltzmann transport that showed scalable work with uniform P$^0$ angle in the streaming and scattering limits. This paper details the practical requirements of using this new iterative method with angular adaptivity. Hence we modify our angular adaptivity to occur directly in P$^0$ space, rather than the Haar space. We then develop a modified stabilisation term for our FEM method that results in scalable growth in the number of non-zeros in the streaming operator with P$^0$ adaptivity. We can therefore combine the use of this iterative method with P$^0$ angular adaptivity to solve problems in both the scattering and streaming limits, with close to fixed work and memory use. 

We also present a CF splitting for multigrid methods based on element agglomeration combined with angular adaptivity, that can produce a semi-coarsening in the streaming limit without access to the matrix entries. The equivalence between our adapted P$^0$ and Haar wavelet spaces also allows us to introduce a robust convergence test for our iterative method when using regular adaptivity. This allows the early termination of the solve in each adapt step, reducing the cost of producing an adapted angular discretisation. 
\end{abstract}
\begin{keyword}
Radiation transport \sep Boltzmann \sep Angular adaptivity \sep Haar wavelets \sep AIRG
\end{keyword}

\end{frontmatter}
\section{Introduction}
\label{sec:Introduction}
In this work we consider the mono-energetic steady-state form of the Boltzmann Transport Equation (BTE) written as
\begin{equation}
\bm{\Omega} \cdot \nabla_{\bm{r}} \psif + \sigma_\textrm{t} \psif - \int_{\bm{\Omega}'} \sigma_\textrm{s} (\bm{r}, \bm{\Omega}' \rightarrow \bm{\Omega}) \psifd \textrm{d}\bm{\Omega}'  = S_{\textrm{e}}(\bm{r}, \bm{\Omega}),
\label{eq:bte}
\end{equation}
where the number of particles moving in direction $\bm{\Omega}$, at spatial position $\bm{r}$ is given by the angular flux, $\psif$. The macroscopic total and scattering cross section are given by $\sigma_\textrm{t}$ and $\sigma_\textrm{s}$, respectively, with an external source of $S_\textrm{e}$. 

Solving \eref{eq:bte} can be difficult given the dimensionality of the problem; previously we presented methods for building adapted angular discretisations for the BTE \cite{Dargaville2019, Dargaville2019a, Dargaville2019b}. This allowed angular resolution to be focused where important in space/energy. \cite{Dargaville2019} used anisotropic adaptivity on the sphere in a Haar wavelet space, which was built on top of an underlying nested P$^0$ space. The angular matrices in Haar wavelet space cannot be formed scalably, as the number of non-zeros (nnzs) grows non-linearly with angular refinement. As such, we built a matrix-free multigrid method to solve the adapted problems, which mapped the solution into the underlying P$^0$ space in $\mathcal{O}(n)$ and performed a matrix-vector product with the P$^0$ angular matrices, which is scalable, before mapping back to Haar space. 

This method resulted in a reduction in time to solve and memory use when compared to uniform discretisations. We showed that our adaptive process was scalable, enabling the use of up to 15 levels of hierarchical refinement in angle; unfortunately in the streaming limit the matrix-free multigrid could not solve the linear systems with fixed work. This is due to the BTE becoming hyperbolic in the limit of no scattering and we deliberately chose not to use Gauss-Seidel/sweep based smoothers, in an effort to build an iterative method that could scale well on unstructured grids in parallel. 

Recently we showed \cite{Dargaville2023} that by solving in P$^0$ space directly, we could build an iterative method without sweeps with excellent performance in the streaming limit of the BTE. The work in this paper is therefore based on building an adapted angular discretisations in P$^0$ space and combining this with our new iterative method, enabling the efficient use of adapted angular discretisations in both streaming and scattering problems. 

Here we present four contributions: the first is a sparsity-controlled adaptive P$^0$ discretisation for the BTE; the second is the use of an adapted P$^0$ discretisation with our AIRG-based iterative method; the third is a method for selecting coarse (C) and fine (F) points for a multigrid hierarchy that can be combined with angular adaptivity to provide semi-coarsenings without requiring matrix entries; and finally a convergence test for iterative methods that decreases the cost of forming an adaptive angular discretisation with regular adaptivity.
\section{Discretisations}
\label{sec:Discretisations}
The spatial and angular discretisations used in this work are based on that we used previously and we describe them below \cite{hughes_variational_1998, hughes_multiscale_2006, candy_subgrid_2008, buchan_inner-element_2010, Dargaville2019, Dargaville2023}.
\subsection{Angular discretisation}
\label{sec:ang_discs}
We use a P$^0$ DG FEM in angle, with the lowest resolution given by one P$^0$ element per octant (similar to an S$_2$ discretisation). Subsequent levels of refinement come from dividing each element in four, at the halfway points of the azimuth angle and cosine of the polar angle. All the elements at a given level of refinement therefore have constant area and we normalise our constant basis functions so the uniform (and hence diagonal) angular mass matrix is the identity. This P$^0$ discretisation is similar to an S$_n$ product quadrature and features elements that cluster around the poles with uniform refinement. This is not a desirable feature for a uniform discretisation of the BTE, but the nested nature of the refinement means it is simple to build an adapted angular discretisation and hence refinement around the poles only occurs if it is required.  

We can build a Haar wavelet discretisation on top of this P$^0$ space, with the hierarchy in the nested elements replaced with a hierarchy in the wavelet basis functions. These two discretisations are exactly equivalent. As in \cite{Dargaville2023} we solve in P$^0$ space as we can form an assembled copy of the streaming/removal matrix that has fixed sparsity with angular refinement (see \secref{sec:sub-grid}). The equivalence between the P$^0$ and Haar spaces is still very useful as it allows us to easily tag which angular elements require refinement/de-refinement and hence it enables the reduced tolerance solves discussed below. 
\begin{figure}[th]
\centering
\subfloat[][Max level = 2]{\label{fig:forward_middle_step_1}\includegraphics[width =0.25\textwidth]{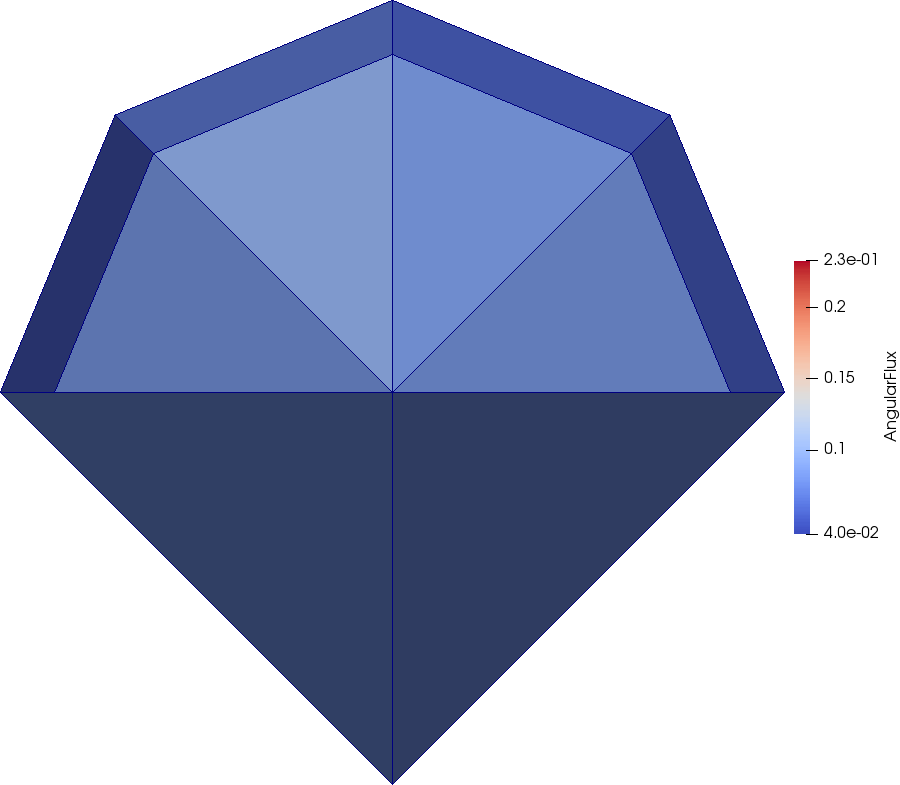}}
\subfloat[][Max level = 3]{\label{fig:forward_middle_step_2}\includegraphics[width =0.25\textwidth]{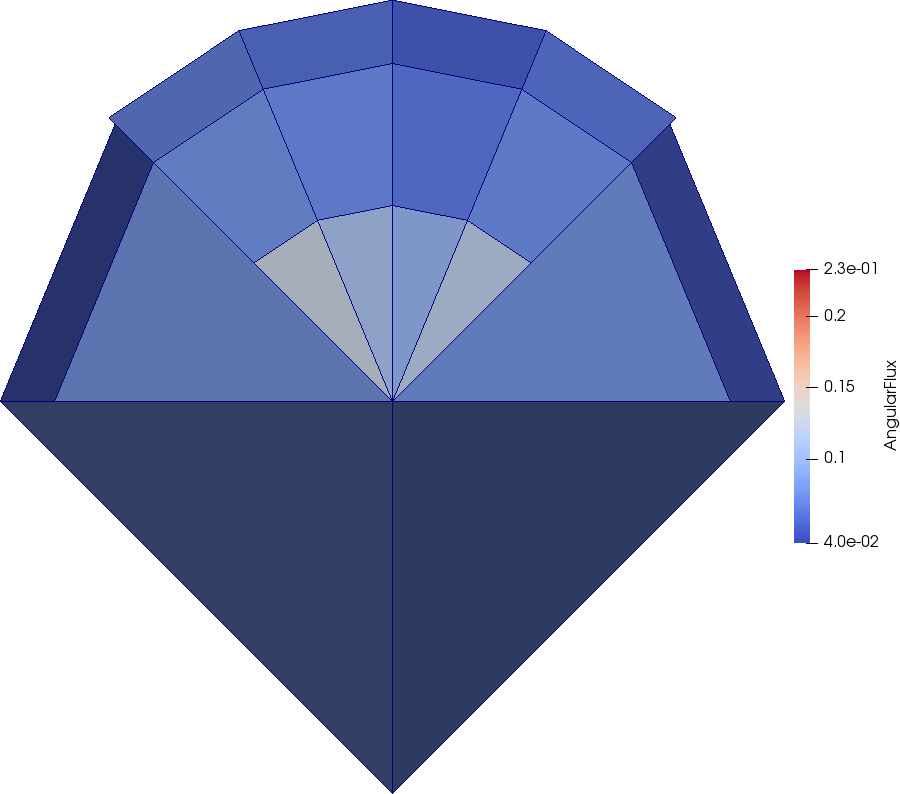}}
\subfloat[][Max level = 4]{\label{fig:forward_middle_step_3}\includegraphics[width =0.25\textwidth]{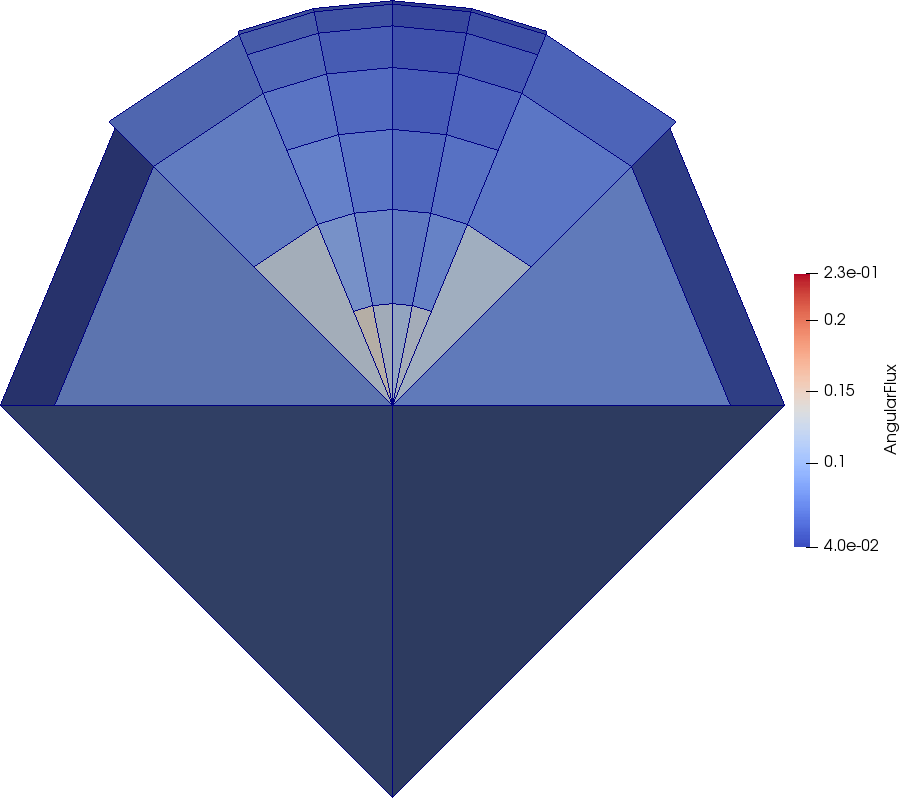}}
\subfloat[][Max level = 5]{\label{fig:forward_middle_step_4}\includegraphics[width =0.25\textwidth]{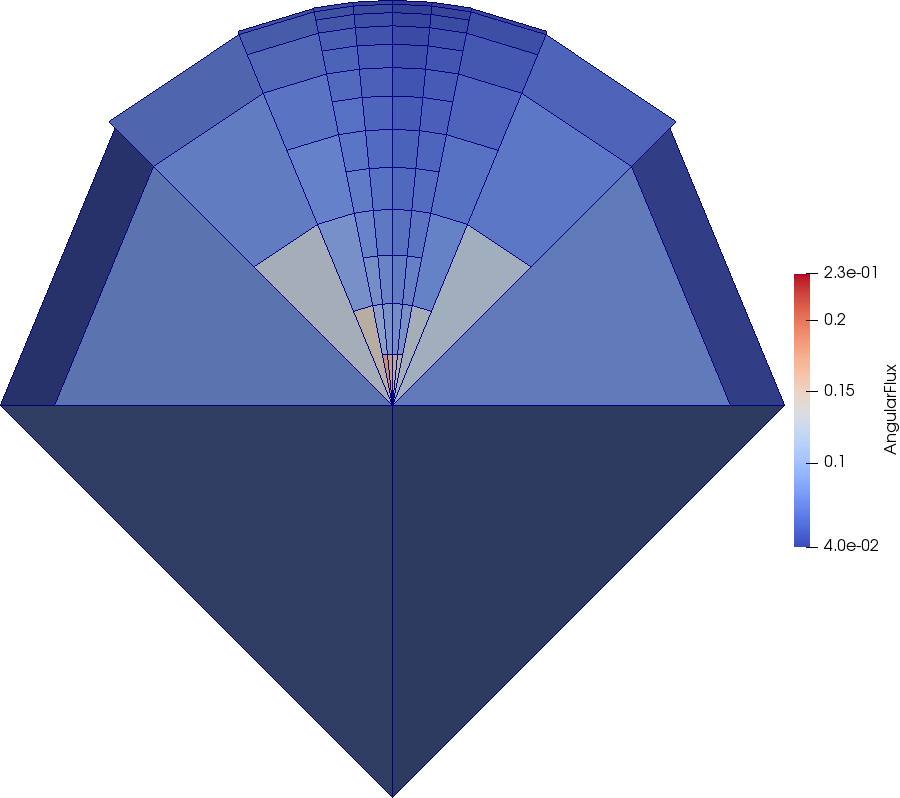}}
\caption{Angular domains showing angular adaptivity focused around a single ``important'' direction, namely $\mu \in [0, 1]$ and $\omega \in [1.47976, 1.661832]$ in a 2D simulation. The P$^0$ angular discretisation is on the $r=1$ sphere, but has been projected onto faceted polyhedra for ease of visualisation. The camera is pointed in the $-z$ direction.}
\label{fig:ang_flux_adapt}
\end{figure}

We noted in \cite{Dargaville2019} that solving in P$^0$ space can easily be used as part of the adaptive process we described previously. \fref{fig:ang_flux_adapt} shows an example of an adapted P$^0$ discretisation at a single node point with several levels of refinement, defined \textit{a priori} in the $+y$ direction. In this work we only show regular adaptivity (based on achieving uniform global accuracy in the solution) given our iterative methods are agnostic to how the adapted discretisations are formed; regular adaptivity is not an optimal method for many problems (particularly in the streaming limit), please see \cite{Dargaville2019, Dargaville2019b} for examples using \textit{a priori} and goal based adaptivity. There have also been a number of other authors who have used adaptivity in angle in Boltzmann applications; these include sparse grid methods \cite{Widmer2008}, S$_n$ methods \cite{Stone2008, Jarrell2010, Jarrell2011, Lau2016,Lau2017, Zhang2018}, FEM methods \cite{Koeze2012, Bennison2014, Murphy2015, Kophazi2015, Hall2017, Favennec2019}, and specifically finite element/wavelet methods \cite{Buchan2006, Buchan2008244, Goffin2015, Goffin2015a, Soucasse2017}. 

Our adaptive process starts with a uniform angular discretisation (typically level 1) and performs a solve in P$^0$ space. The solution is then mapped into Haar wavelet space in $\mathcal{O}(n)$ and an error metric is formed. Regular adaptivity results in a discretisation that minimises the global error in the problem and is simple to perform. In wavelet space, the size of the wavelet coefficients is influenced by both the size of the flux and the smoothness of the underlying function; thresholding the wavelet coefficients with a fixed value is therefore sufficient to drive regular adaptivity. As such, we take the angular flux in Haar space, and scale it by a thresholding coefficient. This coefficient is input by the user and drives refinement as it is decreased (towards a uniform discretisation). We also scale by the maximum of the angular flux, in an attempt to make the thresholding coefficients somewhat problem agnostic. If each resulting wavelet coefficient is greater than 1.0, we tag it for refinement. If it is less than 0.01, we tag it for de-refinement. We therefore know which angular elements in P$^0$ space to refine/de-refine, as they are given by the elements which are in the support of each wavelet; see \cite{Dargaville2019}. This is performed across each spatial point separately. The initial condition for the P$^0$ solve on the adapted discretisation can then be taken from the previous step. The hierarchical nature of the wavelets makes this simple; it is equivalent to interpolating the P$^0$ solution onto the newly adapted P$^0$ discretisation. We then continue this adapted process up to a maximum level of refinement, or number of adapt steps. 

The only difference between this adaptivity process compared with \cite{Dargaville2019} is that we enforce that if a single wavelet is added, all the wavelets on the same level of refinement that share the same support (a maximum of two other wavelets) on the sphere are also added. Similarly, the removal requires that all wavelets with the same support are below the threshold coefficient and all are removed at once. This ensures that when an angular element is refined (or de-refined), we always have 4 P$^0$ elements present, which makes the P$^0$ implementation simpler. In comparison to the matrix-free method in \cite{Dargaville2019}, we must also introduce further controls on the sparsity of our discretisation given we solve in P$^0$ space; this is detailed below.
\subsection{Spatial discretisation}
\label{sec:sub-grid}
Our spatial discretisation is a sub-grid scale FEM, which represents the angular flux as $\bm{\psi} = \bm{\phi} + \bm{\theta}$, where $\bm{\phi}$ is the solution on a ``coarse'' scale and $\bm{\theta}$ is the solution on a ``fine'' scale. We perform a separate finite element expansions on both the fine and coarse scales, with continuous linear basis functions on the coarse scale and discontinuous linear basis functions on the fine scale. We then use (constant) basis functions in angle and enforce that the coarse and fine scales have the same angular expansion on co-located nodes. Our discretised form of \eref{eq:bte} can then be written as
\begin{equation}
\begin{bmatrix}
\mat{A} & \mat{B} \\
\mat{C} & \mat{D} \\
\end{bmatrix}
\begin{bmatrix}
\bm{\Phi} \\
\bm{\Theta} \\
\end{bmatrix}
=
\begin{bmatrix}
\mat{S}_{\bm{\Phi}} \\
\mat{S}_{\bm{\Theta}} \\
\end{bmatrix},
\label{eq:SGS_full}
\end{equation}
The number of unknowns in the coarse scale discretised vector, ${\bm{\Phi}}$, is NCDOFs and the number of unknowns in the fine scale discretised vector, ${\bm{\Theta}}$, is NDDOFs. $\mat{S}_{\bm{\Phi}}$ and $\mat{S}_{\bm{\Theta}}$ are the discretised source terms for both scales. \eref{eq:SGS_full} is built using standard FEM theory and as such $\mat{A}$ and $\mat{D}$ are the standard continuous Galerkin (CG) and discontinuous Galerkin (DG) FEM matrices that result from discretising \eref{eq:bte}. 

A Schur complement of block $\mat{D}$ allows us to solve for the coarse scale variable, ${\bm{\Phi}}$, given by
\begin{equation}
(\mat{A} - \mat{B} \mat{D}^{-1} \mat{C}) {\bm{\Phi}} = \mat{S}_{\bm{\Phi}} - \mat{B} \mat{D}^{-1} \mat{S}_{\bm{\Theta}},
\label{eq:SGS}
\end{equation}
with the fine scale solution then computed with 
\begin{equation}
\bm{\Theta} = \mat{D}^{-1} (\mat{S}_{\bm{\Theta}} - \mat{C} \bm{\Phi}).
\label{eq:theta}
\end{equation}
The addition of the two solutions $\bm{\Psi} = \bm{\Phi} + \bm{\Theta}$ (where the coarse solution $\bm{\Phi}$ has been projected onto the fine space) then gives us our discrete solution. Solving the sub-grid scale equations as posed would be more expensive than solving with a DG FEM discretisation, so we sparsify $\mat{D}$ (see also \cite{buchan_inner-element_2010, Goffin2014, Dargaville_2014, Goffin2015a, Buchan2016, Adigun2018, Dargaville2019, Dargaville2019a}). We replace $\mat{D}^{-1}$ in \eref{eq:SGS} and \eref{eq:theta} with $\hat{\mat{D}}^{-1}$, which is the streaming operator with removal and self-scatter, and vacuum conditions applied on each DG element. This removes the jump terms, resulting in element blocks in $\hat{\mat{D}}$, making the computation of $\hat{\mat{D}}^{-1}$ tractable. With a uniform P$^0$ angular discretisation (as in \cite{Dargaville2023}), this is sufficient to ensure fixed sparsity with either spatial or angular refinement, as $\hat{\mat{D}}$ (and hence $\hat{\mat{D}}^{-1}$) has diagonal blocks. Unfortunately this is not sufficient when we have adapted our P$^0$ angular discretisation with differing angular resolution at each spatial point. If we denote the sparsity of the streaming component $\mat{D}_\Omega$ as $S_\textrm{D} \subset \{(i, j) \, | \, (\mat{D}_\Omega)_{i,j} \neq 0\}$, then we enforce $(\hat{\mat{D}}^{-1})_{i,j} \in S_\textrm{D}$. This is equivalent to using an ILU(0) to invert our sparsified approximation.

\begin{figure}[th]
\centering
\subfloat[][Sparsity of an element matrix from $\mat{D}_\Omega$, which is also enforced on $\hat{\mat{D}}$ and $\hat{\mat{D}}^{-1}$]{\label{fig:d}\includegraphics[width =0.45\textwidth]{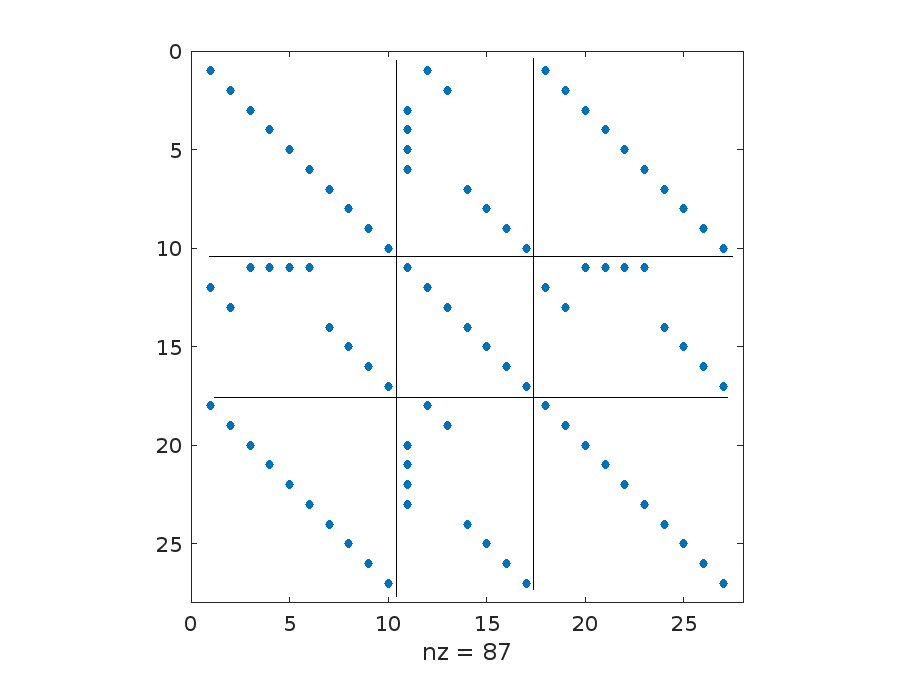}} \hspace{0.1cm}
\subfloat[][Sparsity of an element matrix from $\mat{D}_\Omega^{-1}$]{\label{fig:invd}\includegraphics[width =0.45\textwidth]{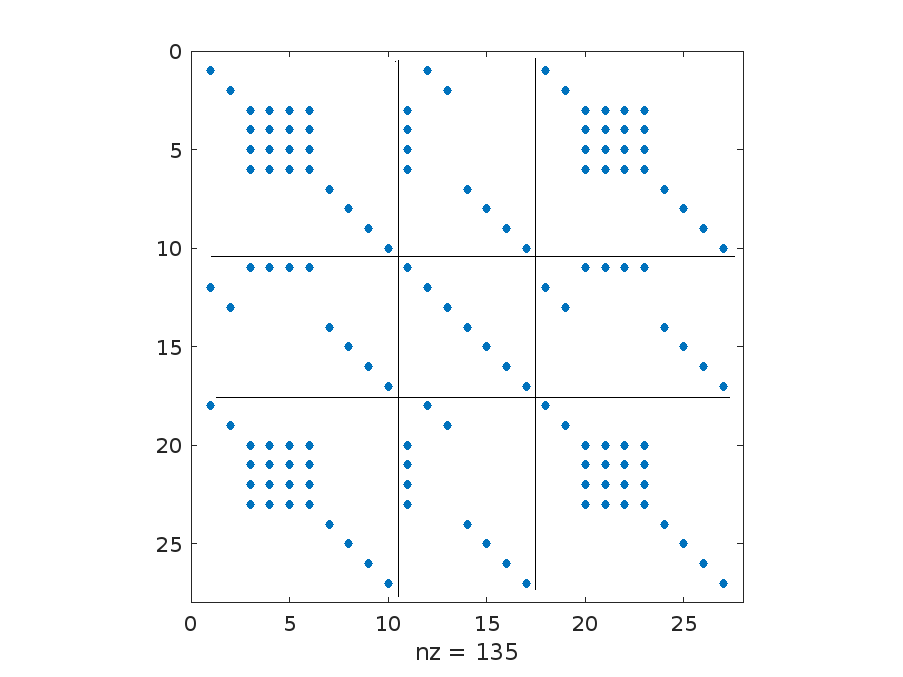}}
\caption{Sparsity of a element streaming matrix from $\mat{D}$ given adapted P$0$ angle in 2D. The three DG nodes have 10, 7 and 10 angles present, respectively; all three nodes start with three octants at level 1 and one octant at level 2, the nodes with 10 angles have one of the level 2 patches further refined to level 3.}
\label{fig:invd_sparse}
\end{figure}

The ILU(0) is only necessary to ensure a fixed sparsity with our adapted P$^0$ space; for example if we have spatial nodes with differently adapted P$^0$ discretisations, the construction of our angular discretisation means that the nnzs in each of the blocks of our element matrices depends on how each adapted angle overlaps the others (i.e., the angular mass matrix is no longer diagonal). \fref{fig:d} shows this on an example element, where the element streaming operator is no longer made up of diagonal blocks. If we inverted this without the sparsity control of an incomplete LU factorisation, we would significantly increase the nnzs in an adapted P$^0$ space, as shown in \fref{fig:invd}; this is in contrast to a uniform P$^0$ space where the nnzs in the inverse is the same as the streaming operator. 

For our adaptivity to be useful, we want the cost of our matvec to scale with the number of \textit{adapted} unknowns. We can imagine a pathological case where, in one dimension for example, the angular resolution varies across the domain such that each (linear) spatial element has one node at the lowest resolution possible and one at the highest. The streaming operator would have the same nnzs as a uniform P$^0$ discretisation at the highest resolution, given the effect of overlapping angular elements in the angular matrices, described above. Thankfully in our experience this pathology does not occur in practice; we show this in \secref{sec:Results}. 
\section{Iterative method}
\label{sec:iterative}
The iterative method we use in this work comes from \cite{Dargaville2023} and is briefly detailed here. We solve a right preconditioned version of \eref{eq:SGS_full} given by 
\begin{equation}
(\mat{A} - \mat{B}\hat{\mat{D}}^{-1} \mat{C}) \mat{M}^{-1} \mat{u} 	 = \mat{S}_{\bm{\Phi}} - \mat{B} \mat{D}^{-1} \mat{S}_{\bm{\Theta}}, \quad \mat{u} = \mat{M} \bm{\psi}.
\label{eq:schur_us_precon}
\end{equation}
We use GMRES(30) to solve this equation and a matrix-free matvec to compute the action of $(\mat{A} - \mat{B}\hat{\mat{D}}^{-1} \mat{C})$. The preconditioner, $\mat{M}^{-1}$, uses the the additive combination of a CG diffusion-synthetic-acceleration (DSA) method and a sparsified form of our streaming/removal operator, which we denote as
\begin{equation}
\mat{M}^{-1} = \mat{M}_\textrm{angle}^{-1} + \mat{M}_\Omega^{-1}.
\label{eq:precon_add}
\end{equation}
The CG DSA preconditioner is based on a CG FEM discretisation of a diffusion equation, $\mat{D}_\textrm{diff}$, with $\mat{R}_\textrm{angle}$ and $\mat{P}_\textrm{angle}$ the restriction/prolongation of the angular flux to the constant moment and hence 
\begin{equation}
\mat{M}_\textrm{angle}^{-1} = \mat{R}_\textrm{angle} \mat{D}_\textrm{diff}^{-1} \mat{P}_\textrm{angle}.
\label{eq:dsa}
\end{equation}

We can rewrite \eref{eq:SGS} as the contribution from a streaming/removal component (denoted with a subscript $\Omega$) and a scattering component (denoted with a subscript S), or 
\begin{equation}
\left(\mat{A}_\Omega - \mat{B}_\Omega \hat{\mat{D}}^{-1} \mat{C}_\Omega \right) {\bm{\Phi}} + \left((\mat{A}_\textrm{S} + \mat{B}_\textrm{S}(y + \hat{\mat{D}}^{-1} \mat{C}_\Omega) + \mat{B}_\Omega y \right) {\bm{\Phi}} = \mat{S}_{\bm{\Phi}} - (\mat{B}_\Omega+ \mat{B}_\textrm{S}) \hat{\mat{D}}^{-1} \mat{S}_{\bm{\Theta}}.
\label{eq:SGS_sep_stream}
\end{equation}
where $y = \hat{\mat{D}}^{-1} \mat{C}_\textrm{S}$ and our fine component is $\bm{\Theta} = \hat{\mat{D}}^{-1} (\mat{S}_{\bm{\Theta}} - (\mat{C}_\Omega + \mat{C}_\textrm{S}) \bm{\Phi})$. In \cite{Dargaville2023} we used $\mat{M}_\Omega^{-1} = (\mat{A}_\Omega - \mat{B}_\Omega \hat{\mat{D}}^{-1} \mat{C}_\Omega)^{-1}$, but here our adapated P$^0$ space in angle requires further sparsification. If we denote the sparsity pattern of $\mat{B}_\Omega$ as $S_\textrm{B} \subset \{(k, l) \, | \, (\mat{B}^e_\Omega)_{k,l} \neq 0\}$, then a sparsified streaming operator is given by 
\begin{equation}
\mat{M}_\Omega^{-1} = \left( \mat{A}_\Omega - (\mat{B}_\Omega (\hat{\mat{D}}^{-1} \mat{C}_\Omega)_{i,j})_{k,l} \right)^{-1}, \quad (i,j) \in S_\textrm{D}, (k,jl) \in S_\textrm{B}.
\label{eq:precon_angle_space}
\end{equation}
With uniform P$^0$ angle, \eref{eq:precon_angle_space} is exactly $\mat{A}_\Omega - \mat{B}_\Omega \hat{\mat{D}}^{-1} \mat{C}_\Omega$; with adapted P$^0$ angle it is equivalent to computing the matrix product $\hat{\mat{D}}^{-1} \mat{C}$ with no fill-in, and then using this result to compute $\mat{B}_\Omega \hat{\mat{D}}^{-1} \mat{C}_\Omega$, again with no fill-in. This is necessary as again the overlap of adapted angular elements means the product of two (element) matrices with sparsity shown in \fref{fig:d} results in the sparsity given in \fref{fig:invd}, which is unacceptable. When computing the action of the matrix triple-product one-by-one as part of a matrix-free matvec for the outer GMRES, the extra non-zeros are not a concern, it is only when we wish to form an assembled version of $\mat{A}_\Omega - \mat{B}_\Omega \hat{\mat{D}}^{-1} \mat{C}_\Omega$ to precondition with that we must take care to ensure no extra fill-in occurs. Practically, we store an assembled copy of \eref{eq:precon_angle_space} to apply the preconditioner so we use this as a replacement for the non-sparsified $\mat{A}_\Omega - \mat{B}_\Omega \hat{\mat{D}}^{-1} \mat{C}_\Omega$ in \eref{eq:SGS_sep_stream}. The difference between these two approaches is simply a modified stabilisation term when we have adapted.

We now require a method to apply the inverses in our preconditioner. In \cite{Dargaville2023} we developed a multigrid method known as AIRG, based on a reduction-style multigrid \cite{Southworth2017, Manteuffel2019}. We used a single V-cycle of AIRG per application of the preconditioner to apply $\mat{M}_\Omega^{-1}$. The diffusion operator was applied with a single V-cycle of \textit{boomerAMG} from \textit{hypre}. We use the same approach here. If we consider a general linear system $\mat{A}\mat{x}=\mat{b}$ we can form a block-system due to a coarse/fine (CF) splitting of the unknowns as 
\begin{equation}
\begin{bmatrix}
\mat{A}_\textrm{ff} & \mat{A}_\textrm{fc} \\
\mat{A}_\textrm{cf} & \mat{A}_\textrm{cc}
\end{bmatrix}
\begin{bmatrix}
\bm{x_\textrm{f}} \\
\bm{x_\textrm{c}}
\end{bmatrix} = 
\begin{bmatrix}
\bm{b_\textrm{f}} \\
\bm{b_\textrm{c}}
\end{bmatrix}.
\label{eq:air_two}
\end{equation}
We discuss the CF splitting further in \secref{sec:F and C point selection}. Ideal prolongation and restriction operators are given by 
\begin{equation}
\mat{P} = 
\begin{bmatrix}
-\hat{\mat{A}}_\textrm{ff}^{-1} \mat{A}_\textrm{fc} \\
\mat{I}
\end{bmatrix}, \quad
\mat{R} = 
\begin{bmatrix}
-\mat{A}_\textrm{cf} \hat{\mat{A}}_\textrm{ff}^{-1} & \mat{I}
\end{bmatrix}, 
\label{eq:prolong}
\end{equation}
with the coarse-grid matrix computed with $\mat{A}_\textrm{coarse}=\mat{R}\mat{A}\mat{P}$. Repeating this process on the coarse-grid then builds a multigrid hierarchy. AIRG forms approximations to $\mat{A}_\textrm{ff}^{-1}$, namely $\hat{\mat{A}}_\textrm{ff}^{-1}$, by using fixed-order GMRES polynomials. This results in a stationary multigrid method that can be applied with just matrix-vector products (i.e., no dot products). One key feature of AIRG is that it doesn't rely on any block or lower-triangular structure in our matrix. This is essential given that our spatial discretisation results in a CG-stencil in \eref{eq:SGS} and hence our linear system does not feature lower-triangular blocks in the streaming limit, unlike a traditional DG FEM.  If we use angular adaptivity in P$^0$ space, different angles in an octant can be coupled across space in the streaming limit, and hence we no longer have independent angle blocks in our matrix. Instead our matrix has at most 4 (in 2D or 8 in 3D) angular blocks given by the octant coupling. We found in \cite{Dargaville2023} that with uniform angle, in both the streaming and scattering limit, our iterative method with AIRG used close to fixed work in both the setup and solve with constant memory consumption. We would hope that the block-independent nature of AIRG gives the same performance when we have adapted in angle; we examine this in \secref{sec:Results}. 

To measure the amount of work required to solve \eref{eq:SGS}, \eref{eq:theta} and form $\bm{\Psi}$, we use the metrics defined in in \cite{Dargaville2023}. This is given in terms of the number of ``Work Units'', which is a FLOP count scaled by the number of FLOPs required to compute a matrix-free matvec of \eref{eq:SGS_sep_stream}. In order to make comparisons with traditional DG FEM/source iteration methods easier, we also present the FLOP count scaled by the number of FLOPs required to compute a matrix-free matvec with a DG FEM discretisation. Please see \cite{Dargaville2023} for a more detailed definition of these quantities. 
\section{CF splitting}
\label{sec:F and C point selection}
In \cite{Dargaville2023} we used the Falgout-CLJP coarsening algorithm implemented in \textit{hypre} to determine the CF splitting required by our AIRG multigrid method (see \cite{Brandt2000} for some general strategies related to CF splittings). In an effort to build cheaper CF splitting algorithms, in this work we also use the element agglomeration algorithms from \cite{Dargaville2019, Dargaville2021}. These methods were vital to our previous work, as we needed to build and apply spatial tables on coarse elements in order to compute matrix-free matvecs on lower spatial grids. In this work we don't require the coarse elements these methods provide, but we can still use the spatial CF points they provide to pick which of our space/angle DOFs are fine and coarse. Given these algorithms only depend on the spatial grid, they can be cheaper to run than algorithms which require access to the matrix entries, though because of this we would expect them to perform less well, given the same spatial CF points would be applied to each angle, and hence ignore the directionality. For a reduction-based multigrid, the selection of ``good'' CF points results in a well-conditioned $\mat{A}_\textrm{ff}$, though as we demonstrate in \secref{sec:Results} this can be ameliorated through the use of strong approximations to $\mat{A}_\textrm{ff}^{-1}$. 

In this paper when using uniform angle (i.e., for the first solve in our adapt process), our element agglomeration CF splitting has no directionality. As mentioned in \secref{sec:ang_discs} however, the P$^0$ angular discretisation used in this work can adapt, refining in ``important'' directions. The adapt process solves a linear system with a uniformly refined angular discretisation in the first step, followed by angular refinement at each spatial point. From there each subsequent adapt steps continues to build anisotropically adapted angular discretisations. For the linear solves after the first step, we can use the directional information contained within the adapted angular discretisations to determine a spatially dependent ``coarsening direction'' and hence a set of CF points with directionality. There are many ways we could define a ``coarsening direction'' at a given spatial point; we could pick the direction of the most refined angular element, compute an average direction on the sphere across all the angular elements, compare the distribution of angles or level of refinement across the sphere, etc. For simplicity in this work, we set the coarsening direction at each spatial node point to be the direction given by the centre of the most refined angular element at that point, or if \textit{a priori} angular refinement is used, the direction of that refinement is used. 

We then have a set of coarsening directions at each spatial node point and this is then used to guide our element agglomeration. The spatial coarse points are the vertices of the coarse agglomerates. If a spatial node is designated coarse, all the angular DOFs on that spatial node are also tagged as coarse. The key to this process is that it must result in a CF splitting that reflects the underlying streaming or scattering limits. One of the benefits of using element agglomeration in this fashion is that after several adapt steps, we could consider the ``coarsening direction'' to be (mostly) converged and hence freeze the element agglomeration. The spatial CF points are then frozen and any additional angular DOFs added in subsequent adapt steps can be cheaply tagged based on the frozen spatial CF points. This could make it cheaper to form the CF splitting with many adapt steps, as we would only need to perform the element agglomeration a small number of times. 

For many problems in the streaming limit, the angular adaptivity will pick out a limited number of key direction(s) and hence the coarsening needs to occur in those directions. We should note that the coarsest level of refinement on our angular domain is given by a single basis function per octant. If we have a single important direction, for example in \fref{fig:ang_flux_adapt} and our coarsening direction has been computed as say, $+y$, there will always be some angles (the unrefined elements) that are not well represented by the coarsening direction. In the limit of angular refinement however, given the AMR-style nested refinement of our angular adaptivity, the majority of angles at that node point will be well represented by the coarsening direction. 

In scattering regions however, the angular adaptivity tends to uniform refinement and hence the coarsening needs to form well-shaped circular/spherical agglomerates (as there is no ``important'' direction), with agglomerates that could have a different number of fine elements compared to the streaming case, given possible differences in optimal coarsening rates. The combination of spatially-dependent coarsening directions formed through angular adaptivity and scattering cross-sections should therefore provide us with the information required to compute a directional element agglomeration algorithm that provides ``good'' CF splitting for the majority of unknowns in our angularly adapted problems. 

There are a few related problems with this scheme. For example, both the linear system in the first adapt step and problems where uniform refinement is triggered far from the scattering limit won't have the required directional information. The first of these is not a major concern as the number of unknowns in the first adapt step is small compared to the number of unknowns at all the other adapt steps. In both cases we must return to using the directionless agglomeration algorithms described in \cite{Dargaville2021}. We can also have the case where we have adapted with the angular DOFs concentrated in small regions of the spatial mesh, but the element agglomeration proceeds with a constant agglomerate size. This can result in many DOFs in those refined regions not being tagged as ``coarse''; the non-uniformity of the distribution of angular DOFs across the spatial mesh therefore persists on the coarse grid. This means our multigrid may have very different operator complexity when compared to uniform angle. These problems may harm the performance of our multigrid methods; we examine this further below but note that the traditional coarsening algorithms that rely on matrix entries can always be used. 
\subsection{Directional element agglomeration method}
\label{sec:Element agglomeration methods}
Previously we presented 7 different element agglomeration methods from the literature and compared their performance on scattering/absorbing problems \cite{Dargaville2021} with the matrix-free multigrid from \cite{Dargaville2019}. Here we present one of those algorithms that has been modified to use directional information; all of the algorithms mentioned in \cite{Dargaville2021} can be modified (and perform similarly).

Our element agglomeration has been modified with very simple heuristics; if the average scattering cross-section is greater than 1 (in this work we use the actual cross-section value on each element; ideally this would just be defined as the mean free path in an element, or an equivalent dimensionless quantity), or the number of angles on a spatial node is ``close'' to that of a uniform discretisation at that level of refinement (we define ``close'' as having $>63\%$ of the uniform angles; this is equivalent to a level 2 discretisation having more than 2 octants refined), then agglomeration proceeds without directional information. If we do proceed with directional coarsening, the tangent vector to the face on the spatial mesh is compared with the average coarsening direction computed on that face (i.e., an average of the coarsening direction across all the spatial nodes on that face). If those two vectors are close to parallel then agglomeration across that face is discouraged; we define close to parallel as the smallest angle between the vectors being less than $\pi/4$. The cleanup routine described in \cite{Dargaville2021} has also been modified to use directional information. Any elements which require cleanup (e.g., unused elements, elements within completely closed agglomerates, etc) are added to the agglomerates closest to the coarsening direction with the smallest number of elements. 

Algorithm 1 uses METIS and requires setting a desired agglomerate size to control the coarsening rate. As mentioned above, typically we would consider the ideal coarsening rate to be dependent on the cross-section. Previously \cite{Dargaville2021} we examined the ideal agglomerate size with our matrix-free multigrid applied to scattering/absorbing problems. The simple grid transfer operators we used in that work meant that the multigrid method could be considered a geometric multigrid and as such we found very large number of elements (20-200 in two and three dimensions, respectively) in an agglomerate were required to keep the grid complexity low and hence maintain good performance. The same is not the case in this work, where our multigrid methods are much closer to AMG/AMGe style methods and we require high quality interpolation/restriction for good performance across all parameter regimes. This requires much smaller numbers of elements in an agglomerate and hence higher grid complexities, although we do use the same coarsening rate in streaming and scattering problems in this work. 

As in \cite{Dargaville2021}, once top grid agglomeration has occurred on an unstructured grid, the lower grids have fundamentally different spatial connectivity and hence come to resemble structured grids, where the agglomerate size can be easily set to a constant across levels. As such we set the agglomerate size to 6 in 2D and 12 in 3D on the top grid, whereas on lower grids we set the agglomerate size to 2 in 2D and 4 in 3D (i.e., we perform aggressive coarsening on the top grid). 
\begin{algorithm}[ht]
\footnotesize
\caption{METIS 5.1 \cite{karypis_metis-unstructured_1995} directional coarsening algorithm}\label{alg:metis}
\begin{algorithmic}[1]
\State Construct the dual-graph of the mesh
\State Construct a coarsening direction, $a$, at every node, by using the centre of the most refined angular element
\State Compute $s$ to be the average scattering cross-section at every node from each of the connected elements
\State Get the number of angles on every node, $n$ and the number of angles with a uniform discretisation, $n_\textrm{uniform}$
\If{$s>1$ or $n \approx n_\textrm{uniform}$}
\State{Set $a=\bm{0}$ on a node}
\EndIf
\State Build integer area/volume weights from the dual-graph by scaling relative face areas and volumes to between 1 and 100
\While {we haven't visited every face - denote current face as $f$}
\State Compute $a_\textrm{avg}$ to be the average coarsening direction on $f$ from each of the vertices on this face
\State Compute $\vartheta$ to be the dot product between a vector parallel to $f$ and $a_\textrm{avg}$
\If{$\vartheta < \pi/4$ and $a_\textrm{avg}\neq\bm{0}$}
\State Set the integer area weight on this face to be 1
\EndIf
\EndWhile
\State Set desired agglomerate size, $s$
\State Set the number of partitions as: int(number of elements / $s$)
\If{number of partitions $>$ 8}
\State Call METIS\_PartGraphKway
\Else
\State Call METIS\_PartGraphRecursive
\EndIf
\State Cleanup the coarsening
\end{algorithmic}
\end{algorithm}
\section{Results}
\label{sec:Results}
Outlined below are two examples problems taken from \cite{Dargaville2023}, in the streaming and scattering limit that we use to test our P$^0$ angular adaptivity. We solve our linear systems with GMRES(30) to a relative tolerance of 1\xten{-10}, with an absolute tolerance of 1\xten{-50} and use an initial guess of zero unless otherwise stated. We use the additive preconditioners defined in \cite{Dargaville2023} and all the same parameters as that work. This is the case even when we have used angular adaptivity and hence may have non-zero initial guesses from previous adapt steps. We do this to ensure fair comparisons across different material regimes at different levels of angular refinement. This is in order to show that the convergence of our method is not dependent on a ``good'' initial condition in some material regimes. In the scattering limit the non-zero guesses from coarser angular discretisations are helpful, but in the streaming limit ray-effects mean that coarse angular discretisations often do not provide good approximations to refined angular discretisations. If using the element agglomeration CF splitting, we rerun this at each adapt step, rather than freeze it after a set number of steps. All tabulations of memory used are scaled by the total NDOFs in $\bm{\psi}$ in \eref{eq:SGS_full}, i.e., NDOFs=NCDOFs + NDDOFs. 
\subsection{Angular adaptivity}
\label{sec:adapt}
\secref{sec:ang_discs} described how our P$^0$ angular discretisation can adapt anisotropically on the sphere, allowing different angular refinement at different spatial points. As such we examine the use of AIRG and our additive iterative method in these adapted systems, along with the directional element agglomeration method described in \secref{sec:Element agglomeration methods}. To see the performance on the same problems with uniform angle, please see \cite{Dargaville2023}. In particular, \cite{Dargaville2023} showed that with uniform angular refinement, in both the streaming and scattering limit we can solve our problem with close to fixed work. The first step of our adapt solves at a (low) uniform level of angular refinement and then this information (through some error metric) is used to trigger one level of refinement where required, followed by subsequent solves/refinements. We therefore expect our subsequent solves to be cheaper than a uniform equivalent given they should have fewer DOFs, and given the sparsity control in \secref{sec:sub-grid} there should be fewer nnzs in our matrices.
\subsubsection{Pure streaming problem}
\label{sec:pure_stream_adapt}
\begin{figure}[h]
\centering
\subfloat[][Scalar flux]{\label{fig:solution_adapt_1}\includegraphics[width =0.25\textwidth]{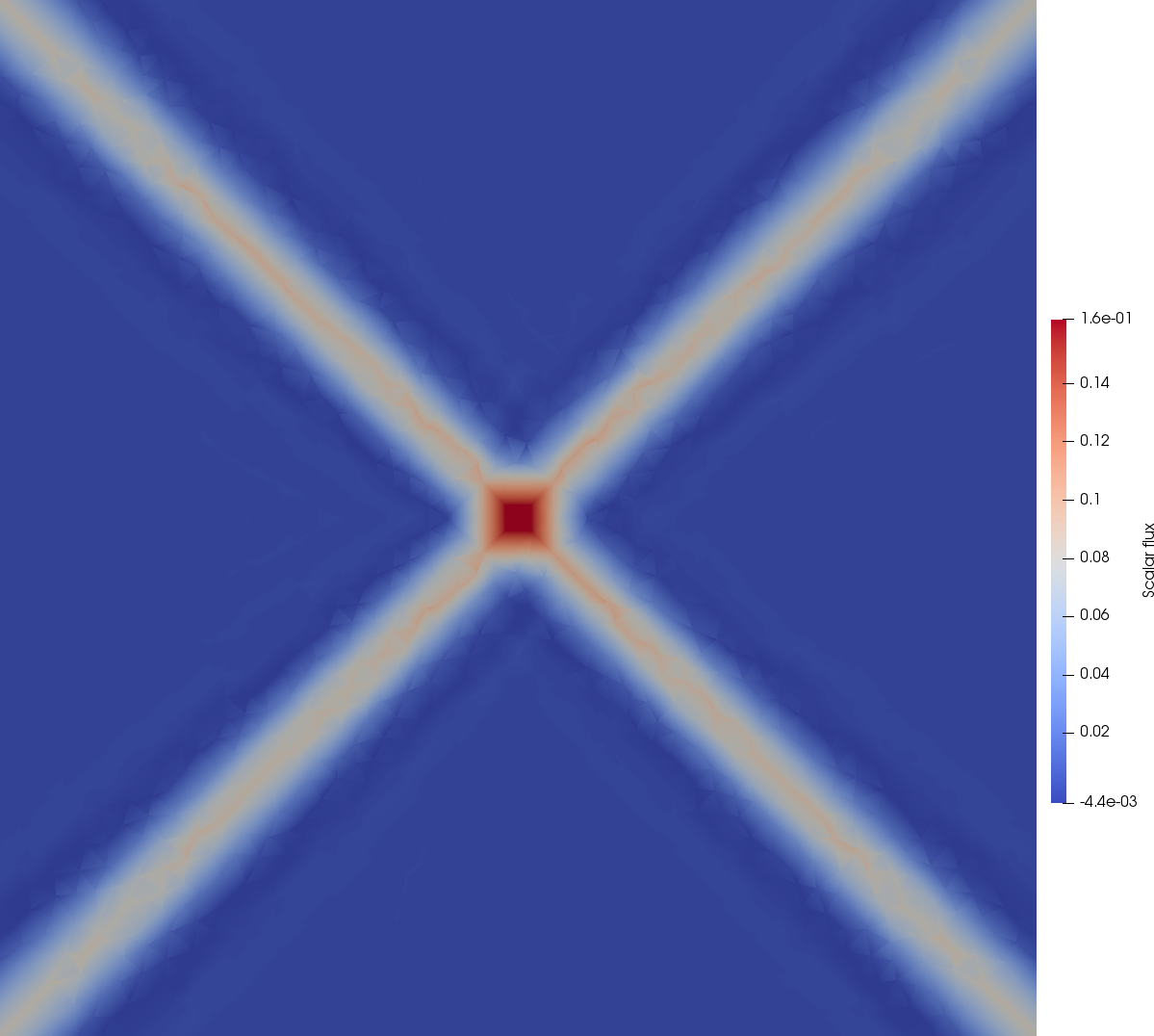}} \quad
\subfloat[][Number of angles across space]{\label{fig:no_angles_1}\includegraphics[width =0.25\textwidth]{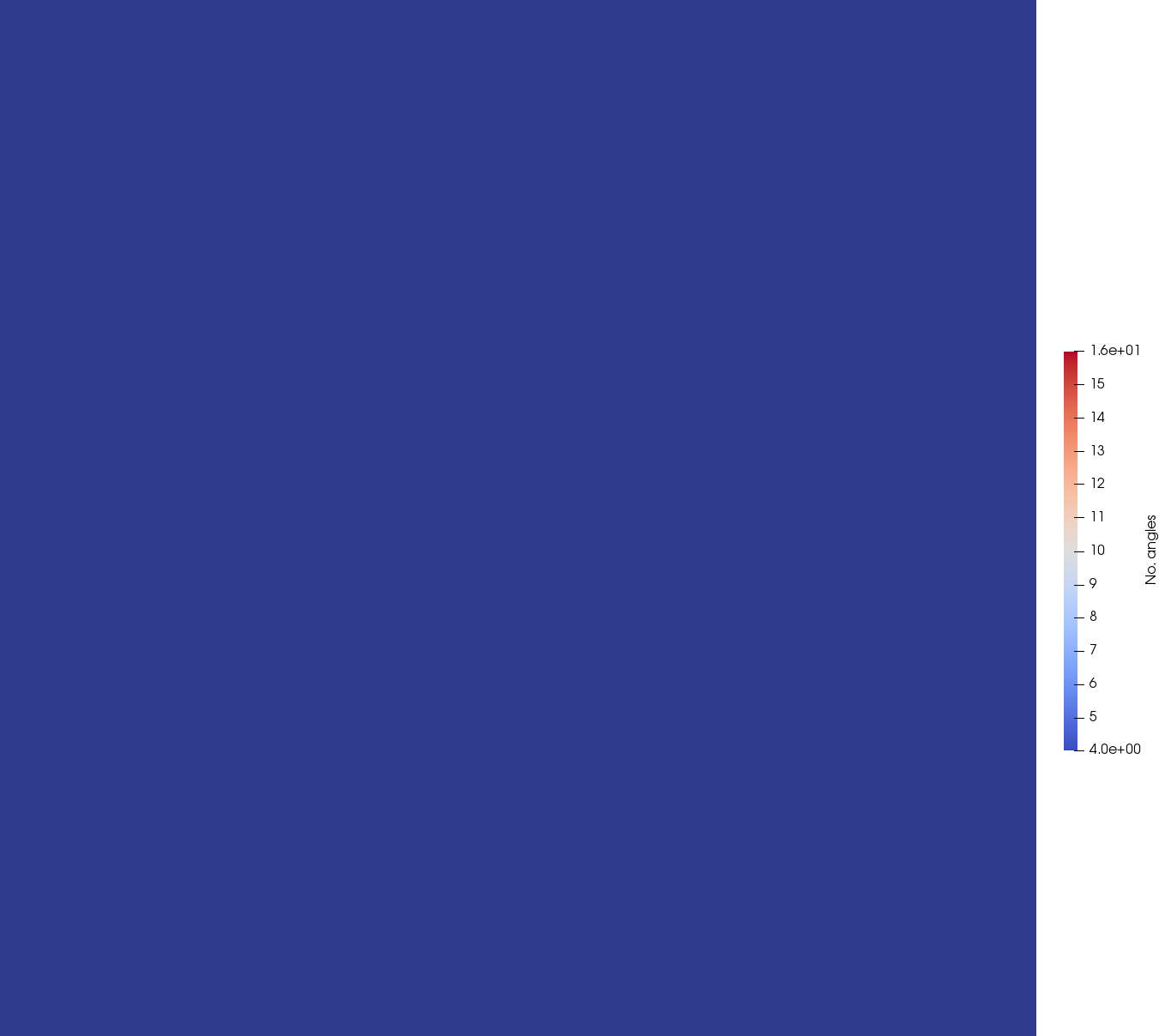}}
\subfloat[][Element agglomeration on the second multigrid level]{\label{fig:coarsening_1} \quad\includegraphics[width =0.225\textwidth]{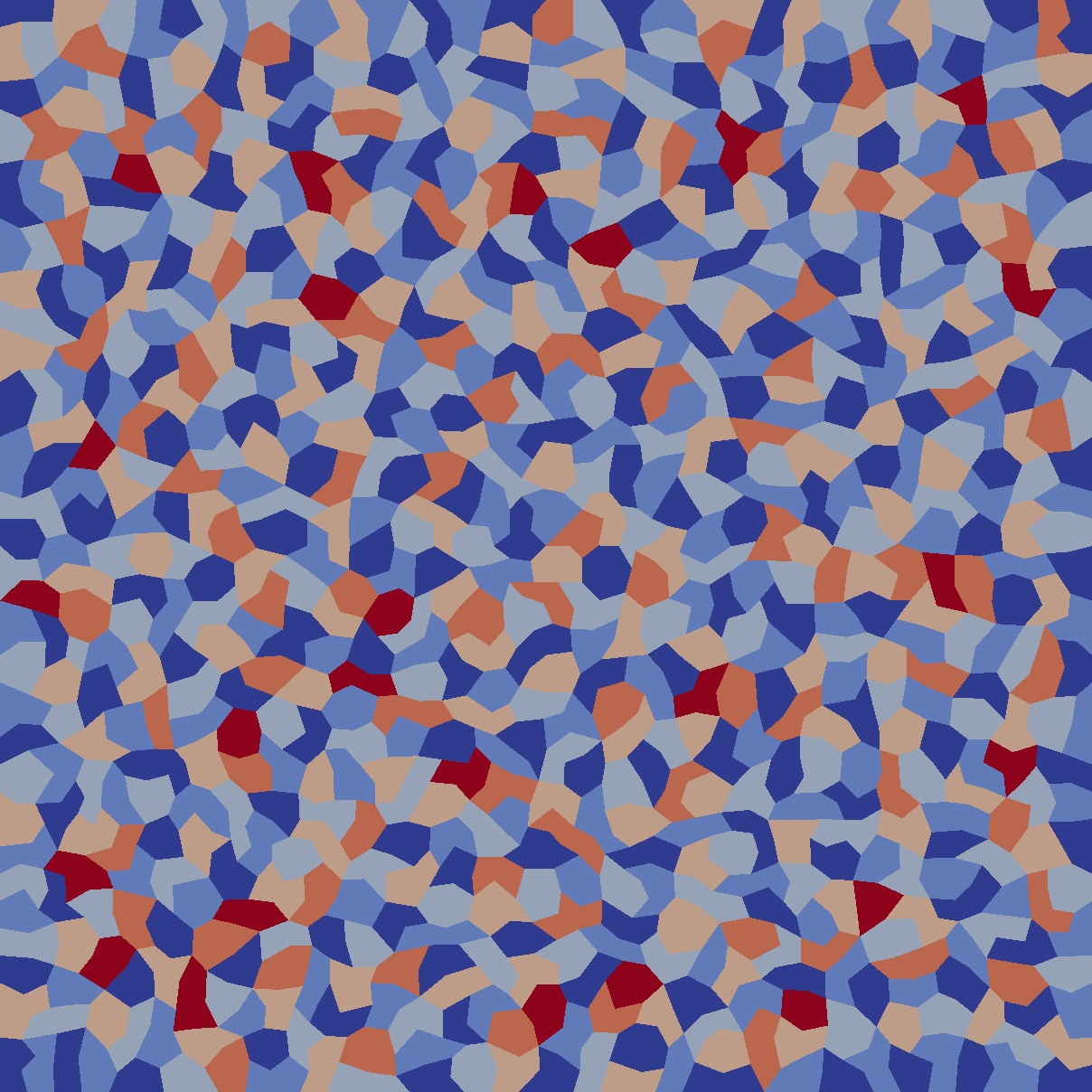}}\\
\subfloat[][]{\label{fig:solution_adapt_2}\includegraphics[width =0.25\textwidth]{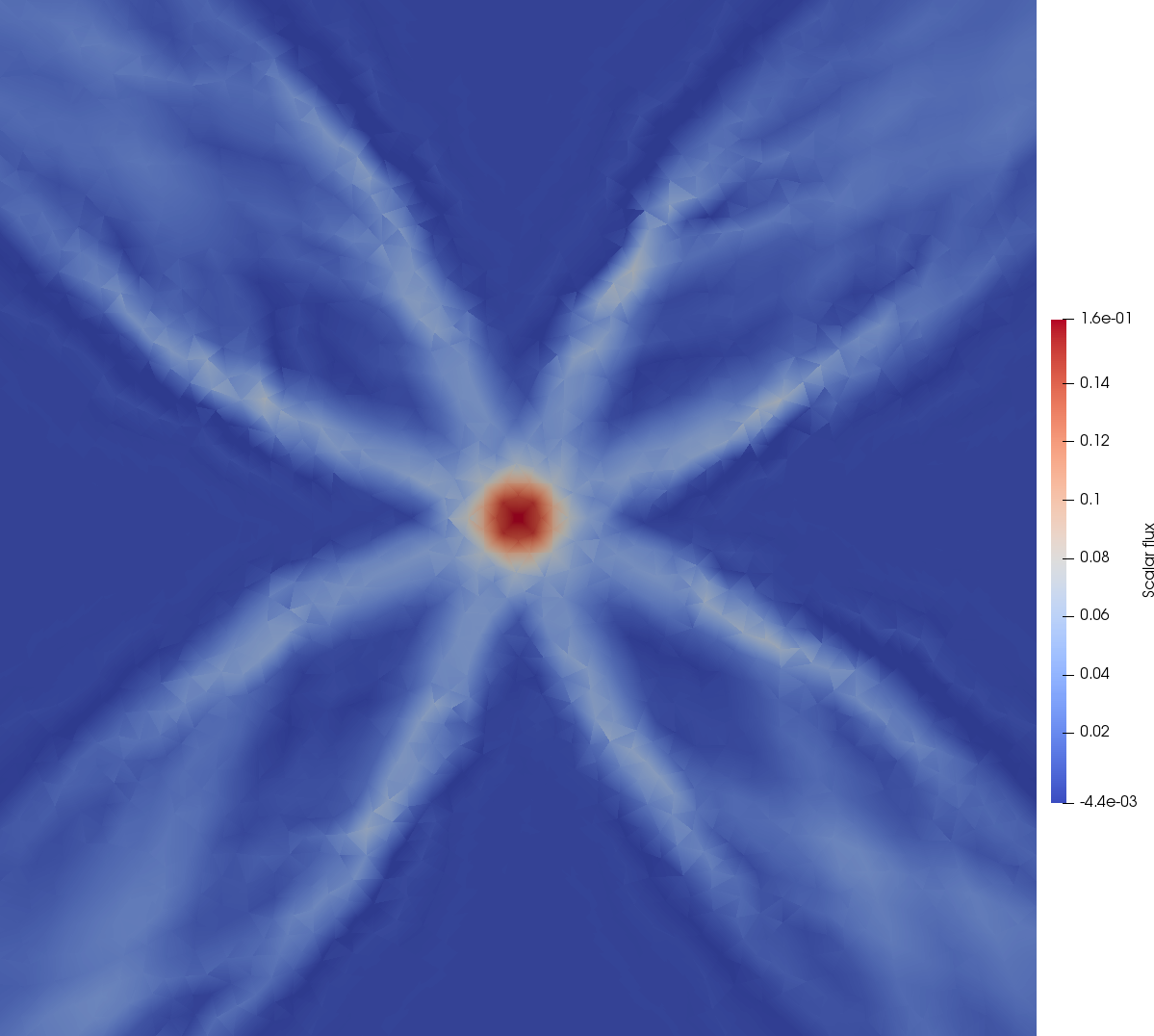}} \quad
\subfloat[][The green dots correspond to the angular flux output in \fref{fig:ang_flux_adapt_2}]{\label{fig:no_angles_2}\includegraphics[width =0.25\textwidth]{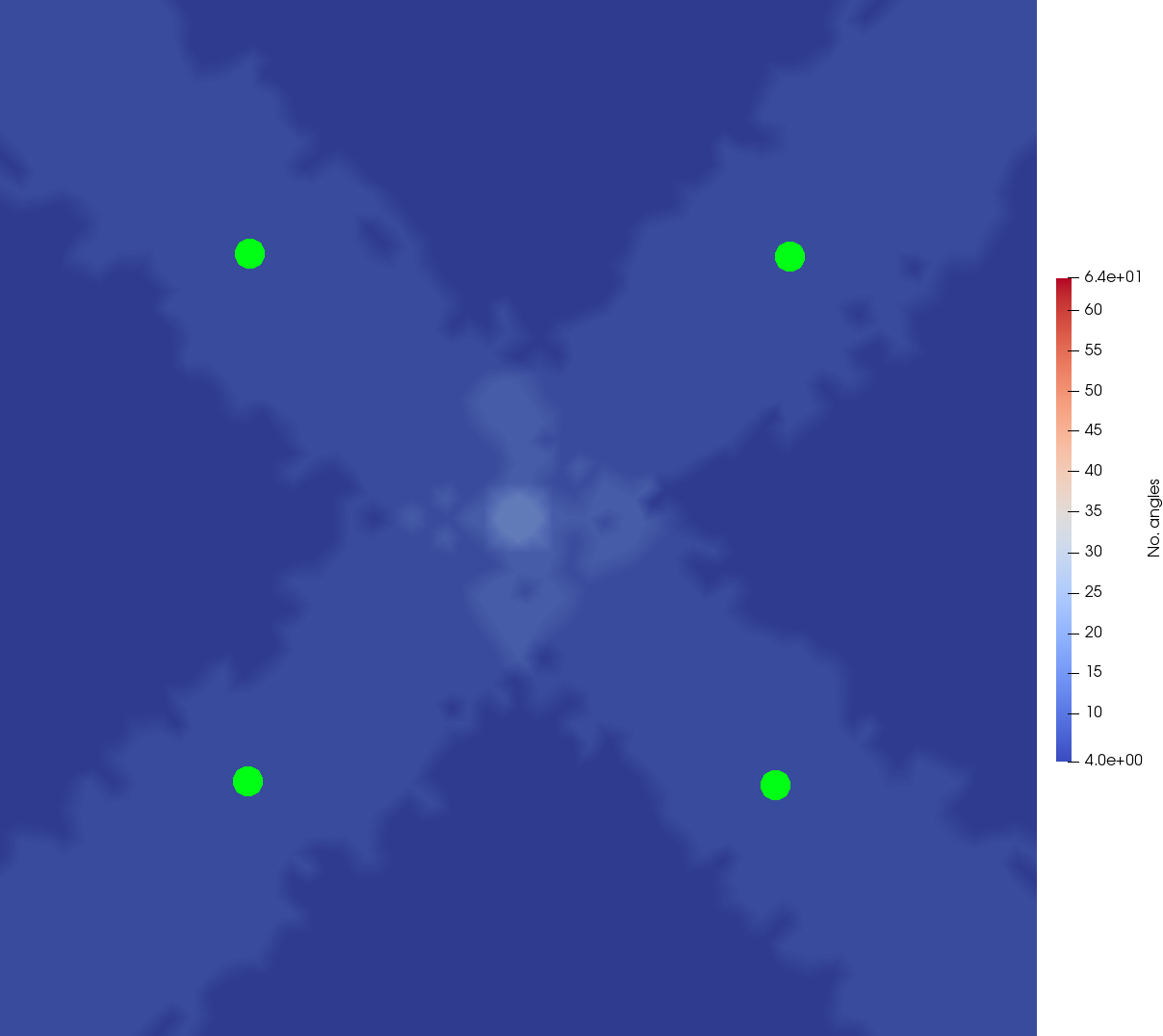}}
\subfloat[][]{\label{fig:coarsening_2} \quad\includegraphics[width =0.225\textwidth]{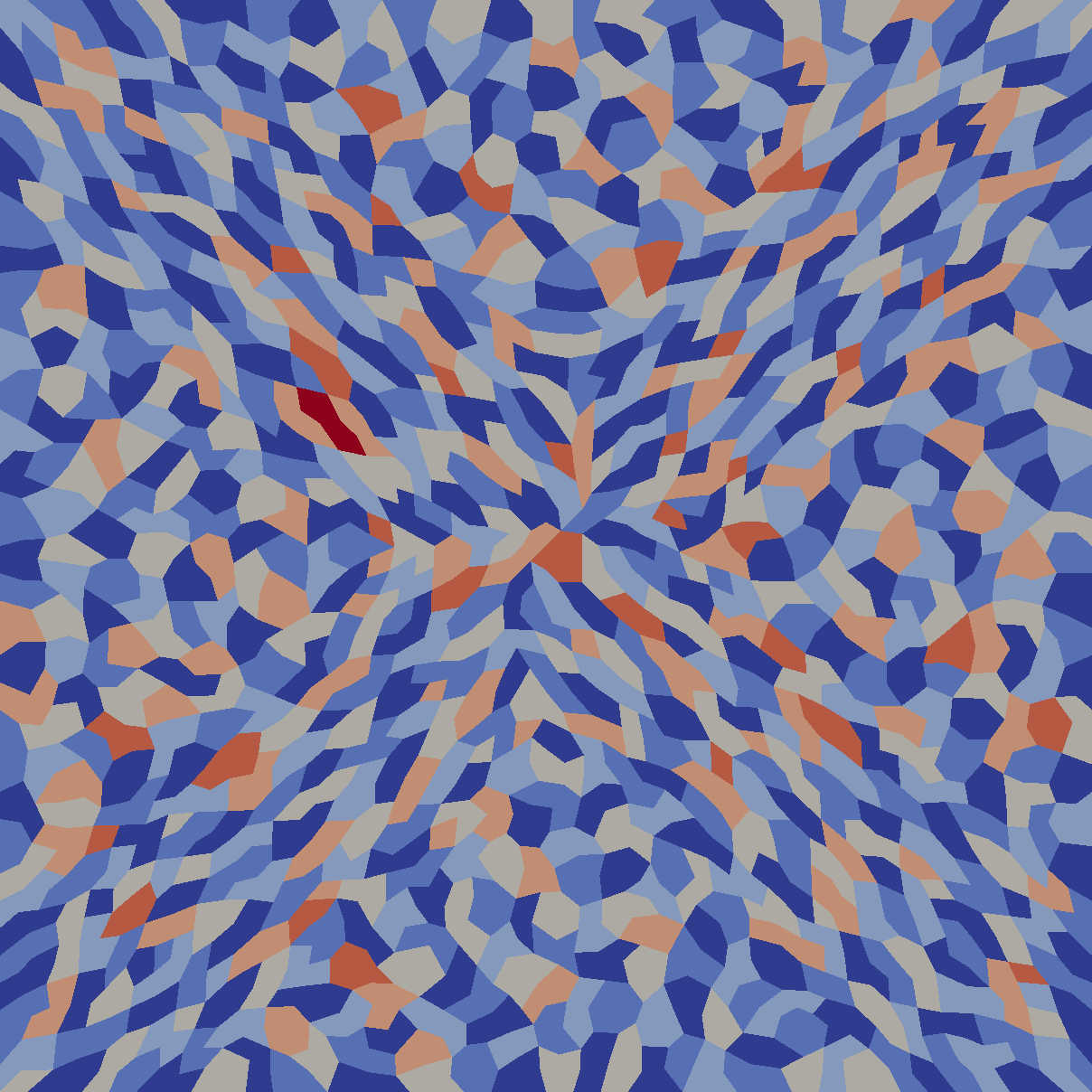}}\\
\subfloat[][]{\label{fig:solution_adapt_3}\includegraphics[width =0.25\textwidth]{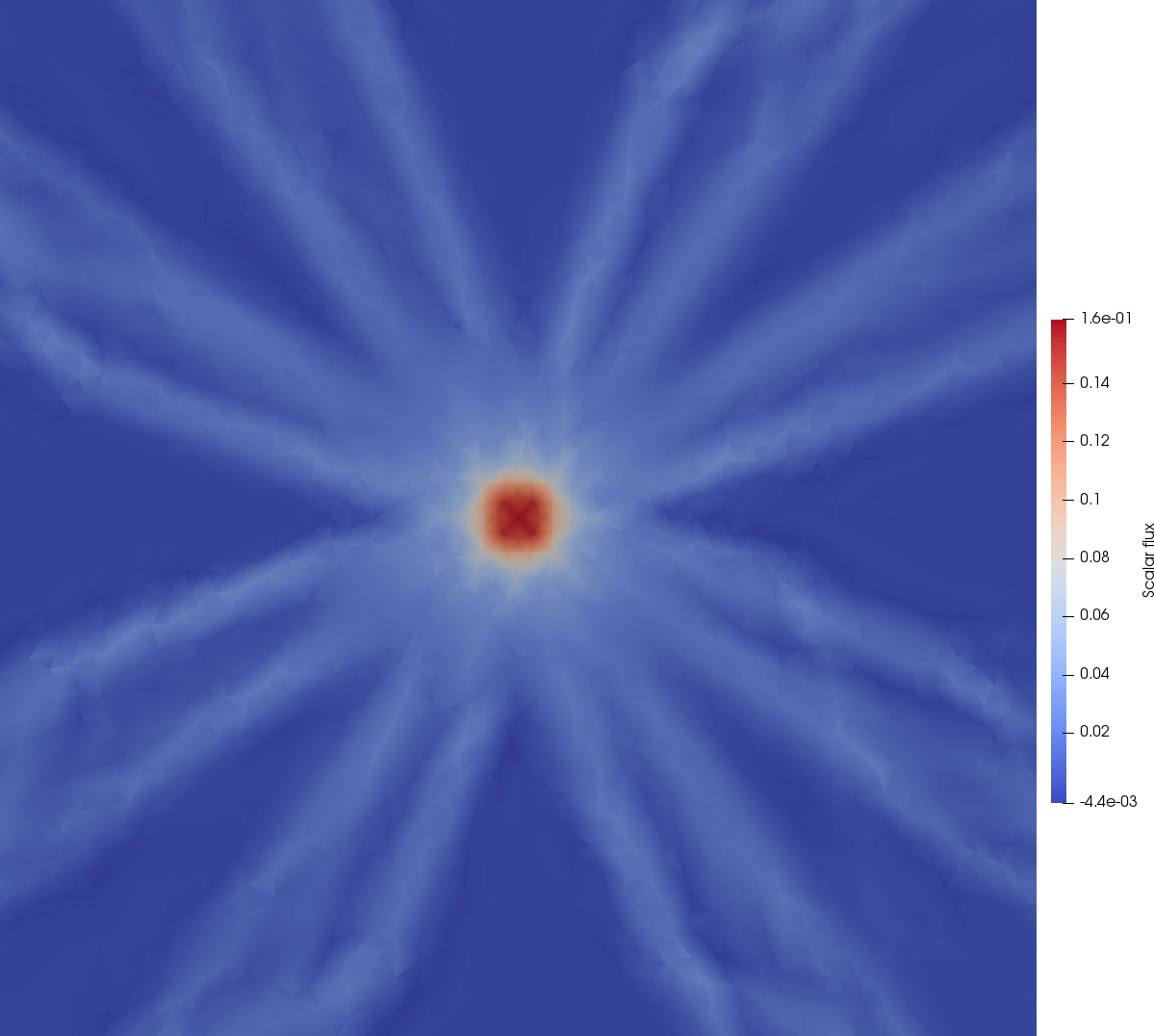}} \quad
\subfloat[][]{\label{fig:no_angles_3}\includegraphics[width =0.25\textwidth]{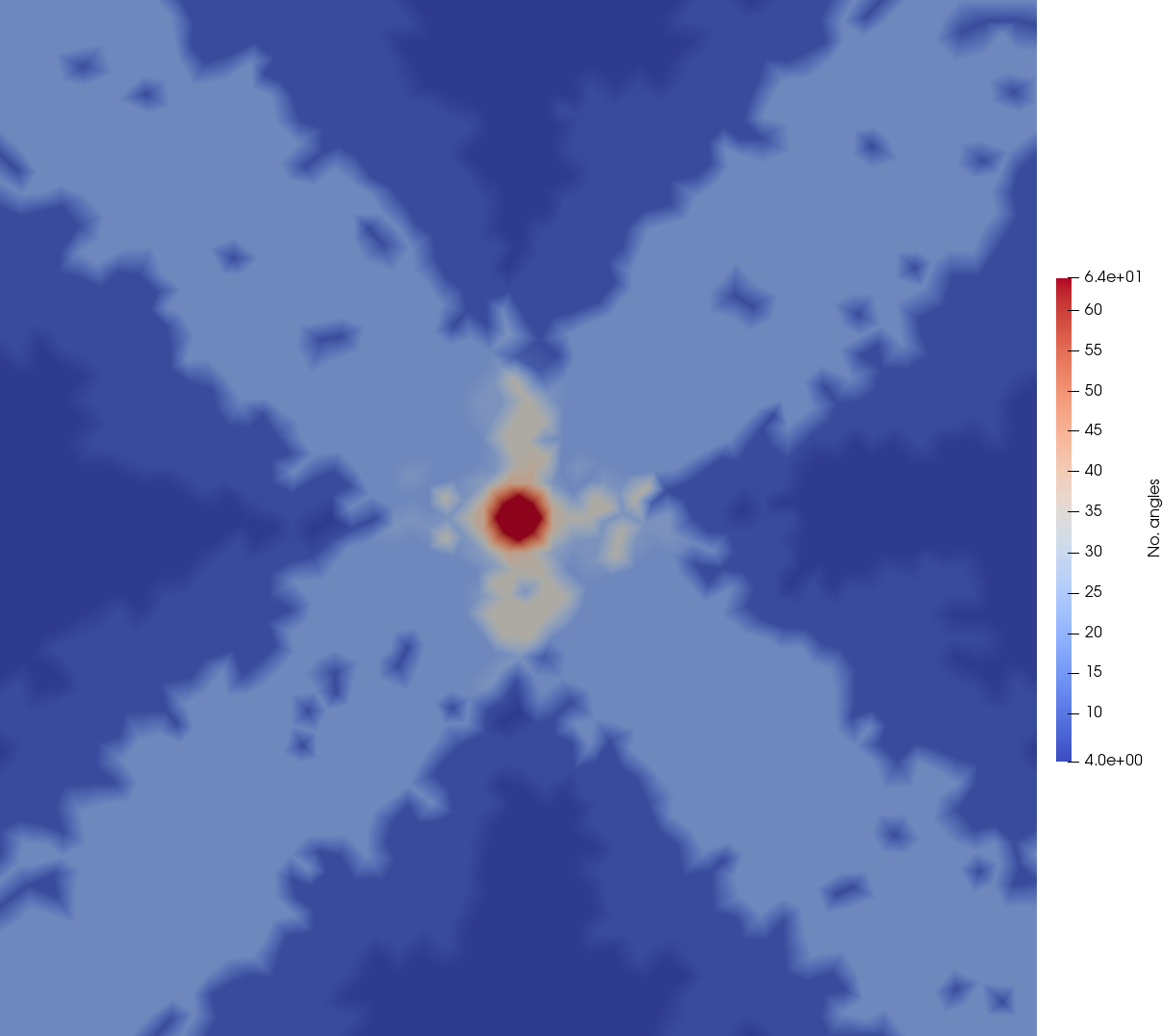}}
\subfloat[][]{\label{fig:coarsening_3} \quad\includegraphics[width =0.225\textwidth]{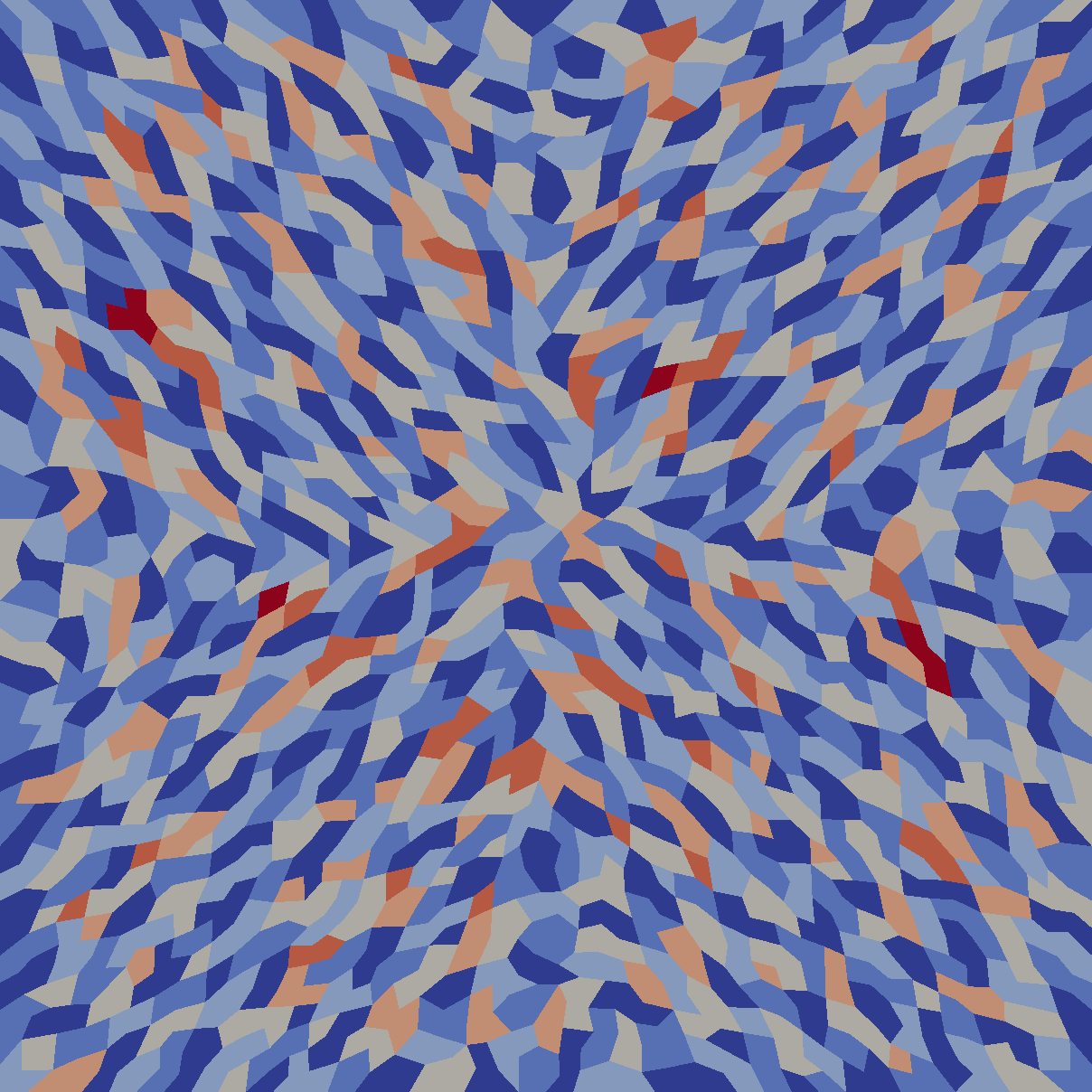}}\\
\subfloat[][]{\label{fig:solution_adapt_4}\includegraphics[width =0.25\textwidth]{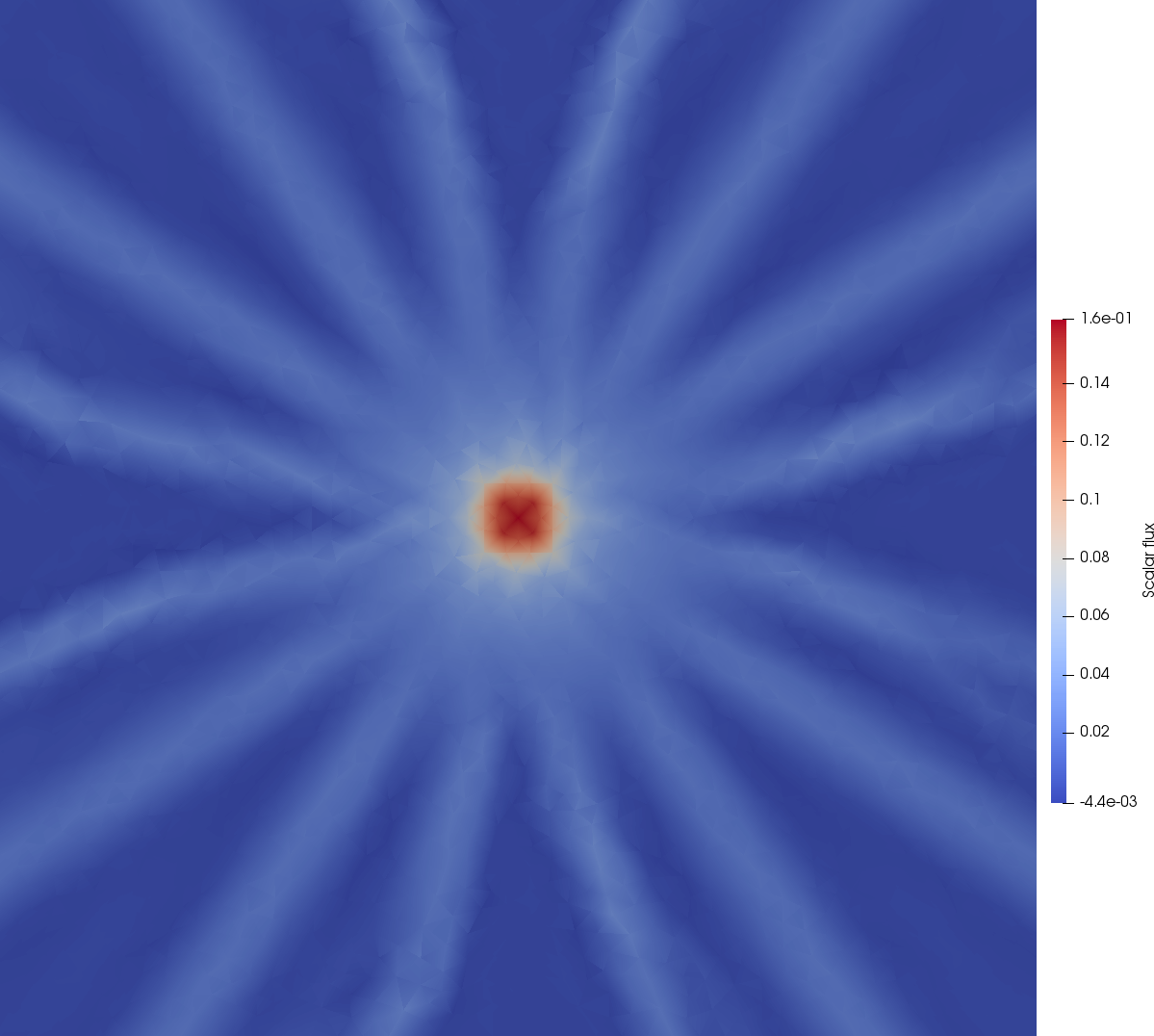}} \quad
\subfloat[][]{\label{fig:no_angles_4}\includegraphics[width =0.25\textwidth]{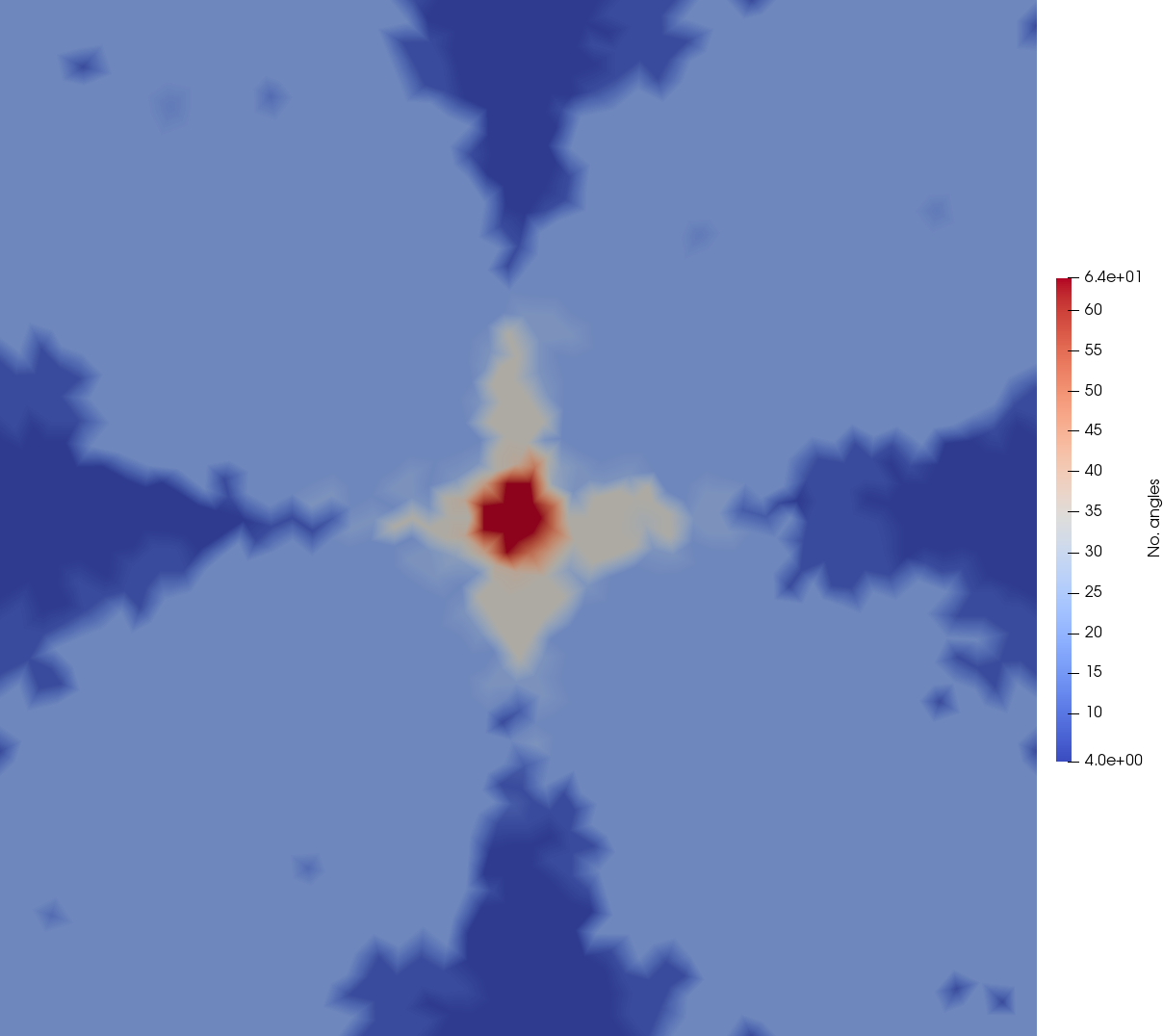}}
\subfloat[][]{\label{fig:coarsening_4} \quad\includegraphics[width =0.225\textwidth]{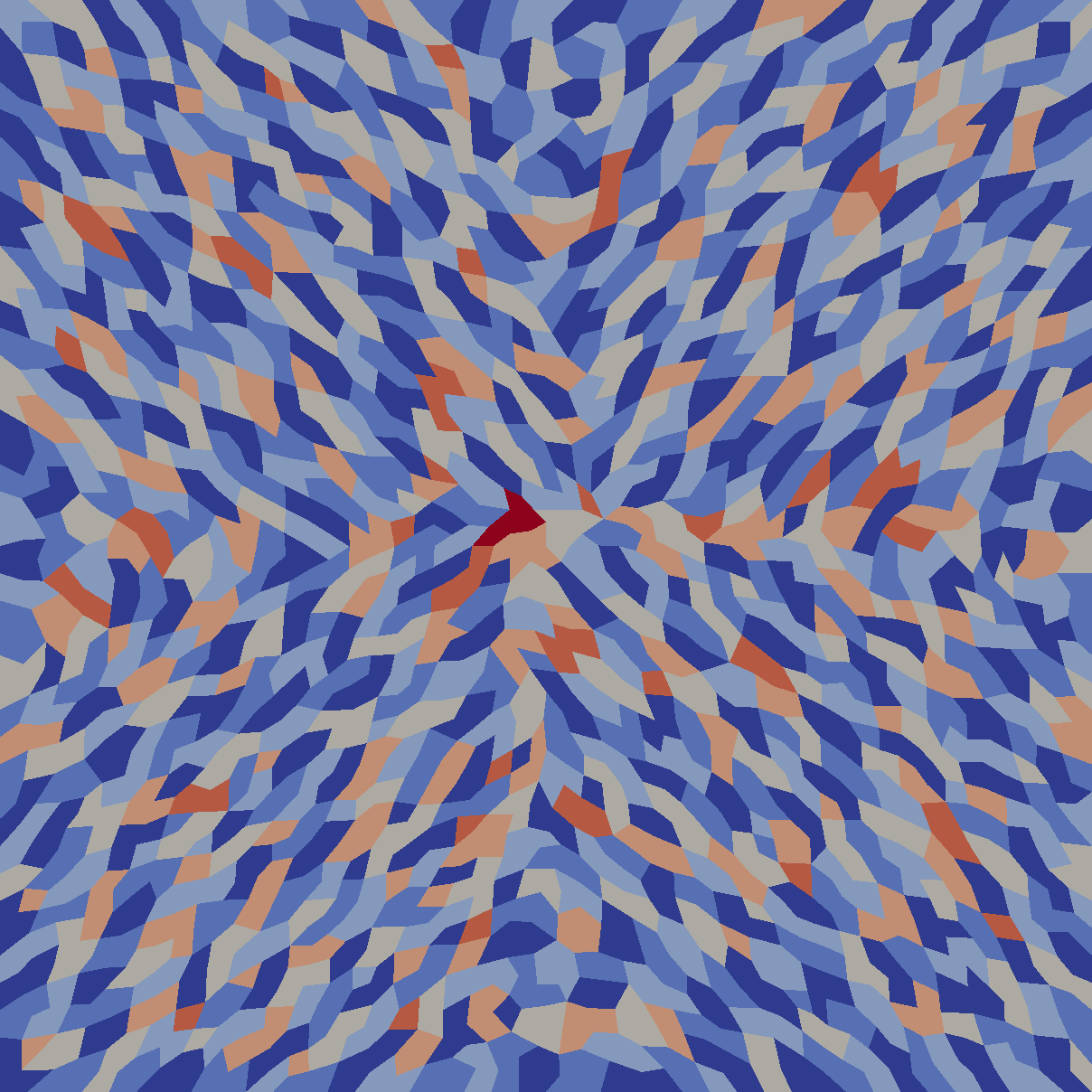}}\\
\subfloat[][]{\label{fig:solution_adapt_5}\includegraphics[width =0.25\textwidth]{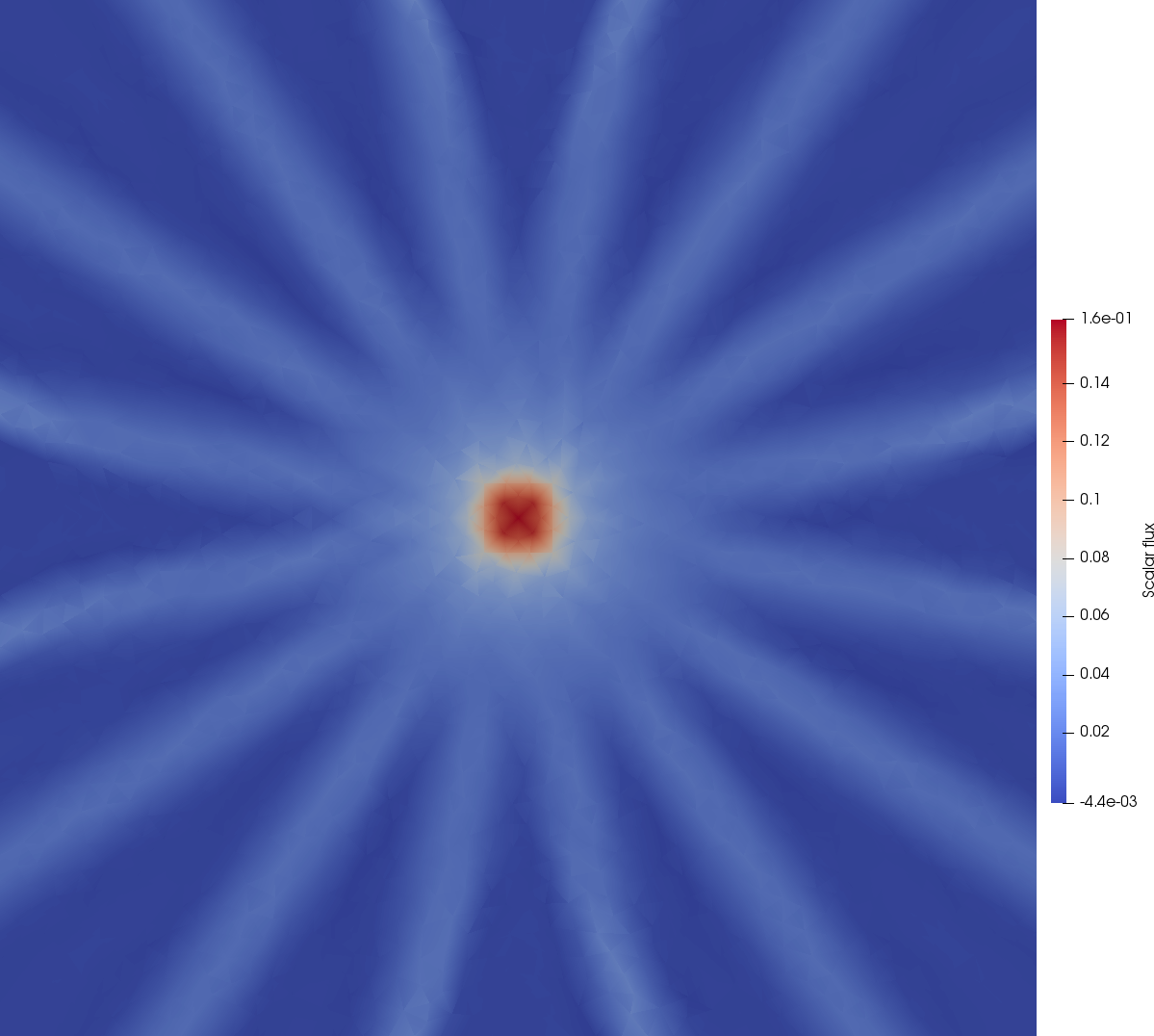}} \quad
\subfloat[][]{\label{fig:no_angles_5}\includegraphics[width =0.25\textwidth]{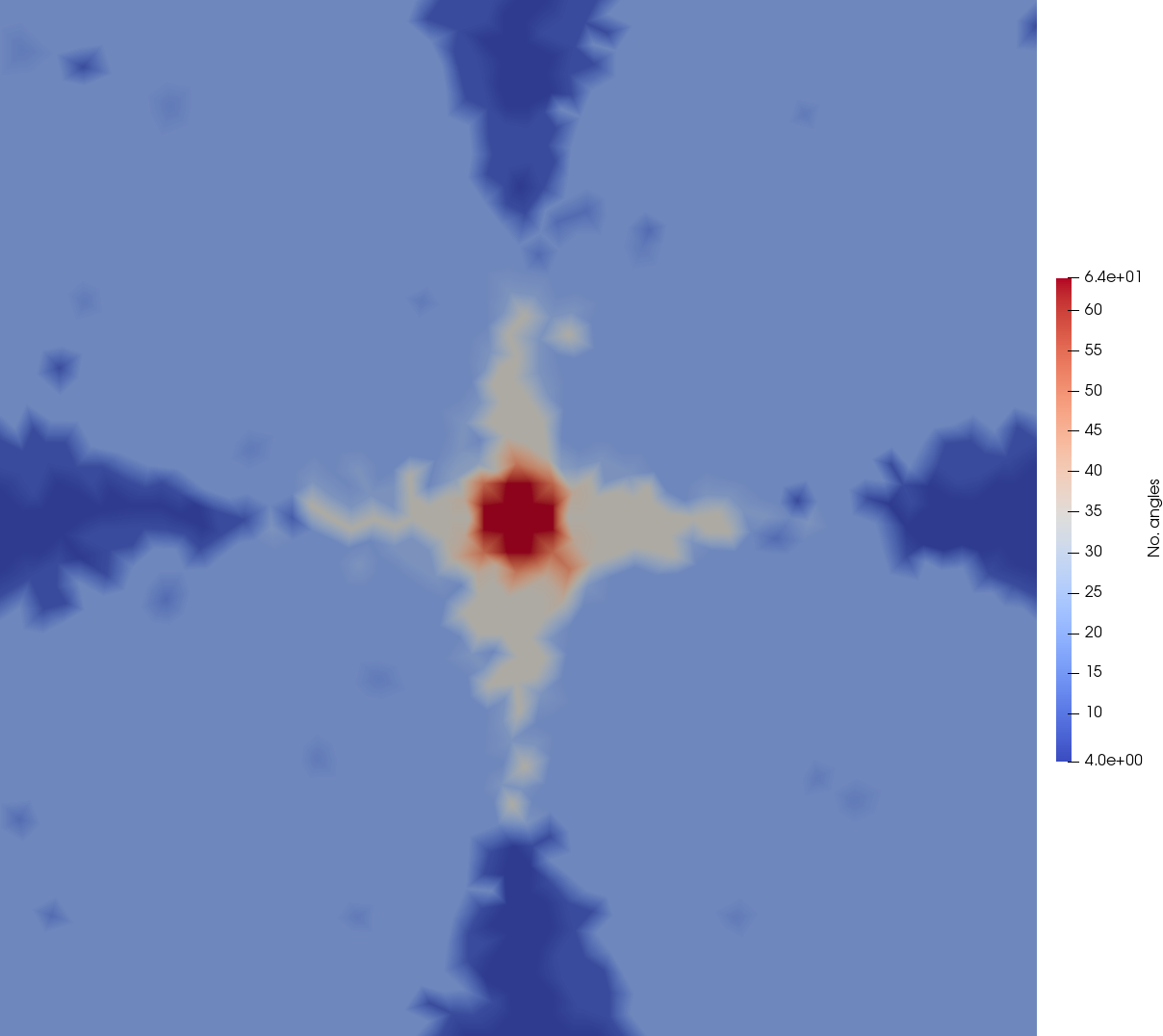}}
\subfloat[][]{\label{fig:coarsening_5} \quad\includegraphics[width =0.225\textwidth]{coarsening_4.png}}\\
\caption{Adapt results for a 2D pure streaming problem with a small source at the centre of the domain, allowing a maximum of 3 levels of regular angular refinement (giving a max. number of angles as 64). The first column shows the scalar flux and the second the number of angles across space. The third column shows the resulting element agglomeration on the second spatial grid if the directional algorithm in \secref{sec:Element agglomeration methods} is used. The rows show consecutive regular adapt steps}
\label{fig:directional_coarsening}
\end{figure}

\begin{figure}[ht]
\centering
\subfloat[][$x=0.75, y=2.25$]{\label{fig:top_left}\includegraphics[width =0.25\textwidth]{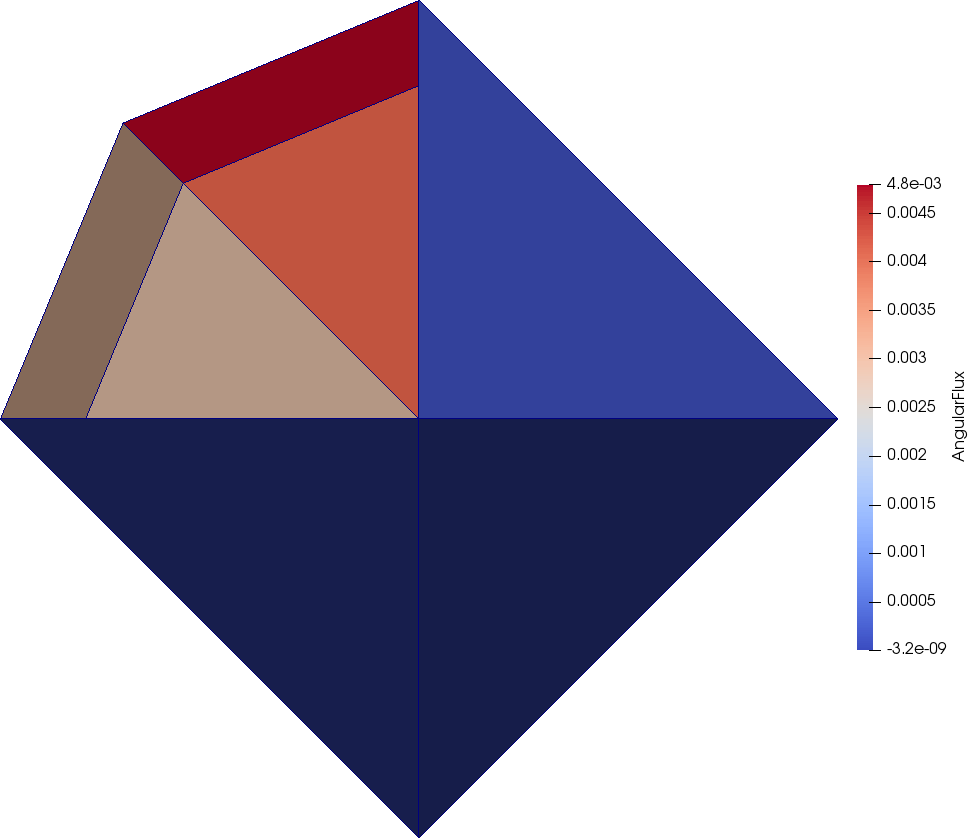}} \quad
\subfloat[][$x=2.25, y=2.25$]{\label{fig:top_right}\includegraphics[width =0.25\textwidth]{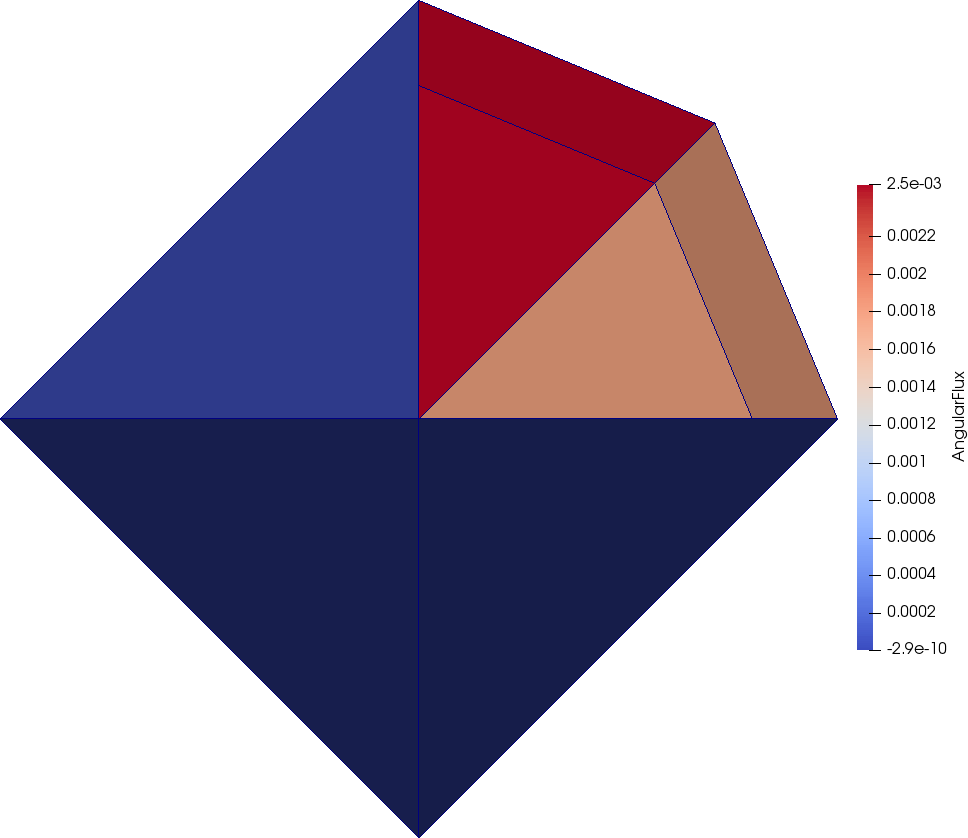}}\\
\subfloat[][$x=0.75, y=0.75$]{\label{fig:bottom_left} \quad\includegraphics[width =0.225\textwidth]{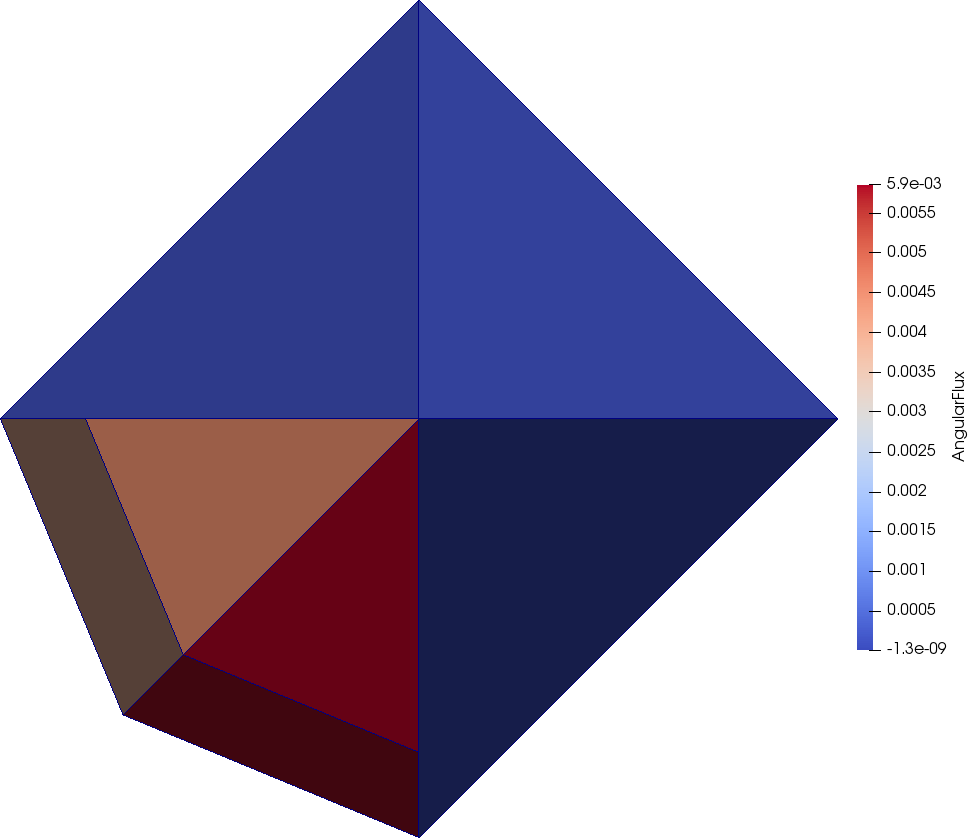}}
\subfloat[][$x=2.25, y=0.75$]{\label{fig:bottom_right} \quad\includegraphics[width =0.225\textwidth]{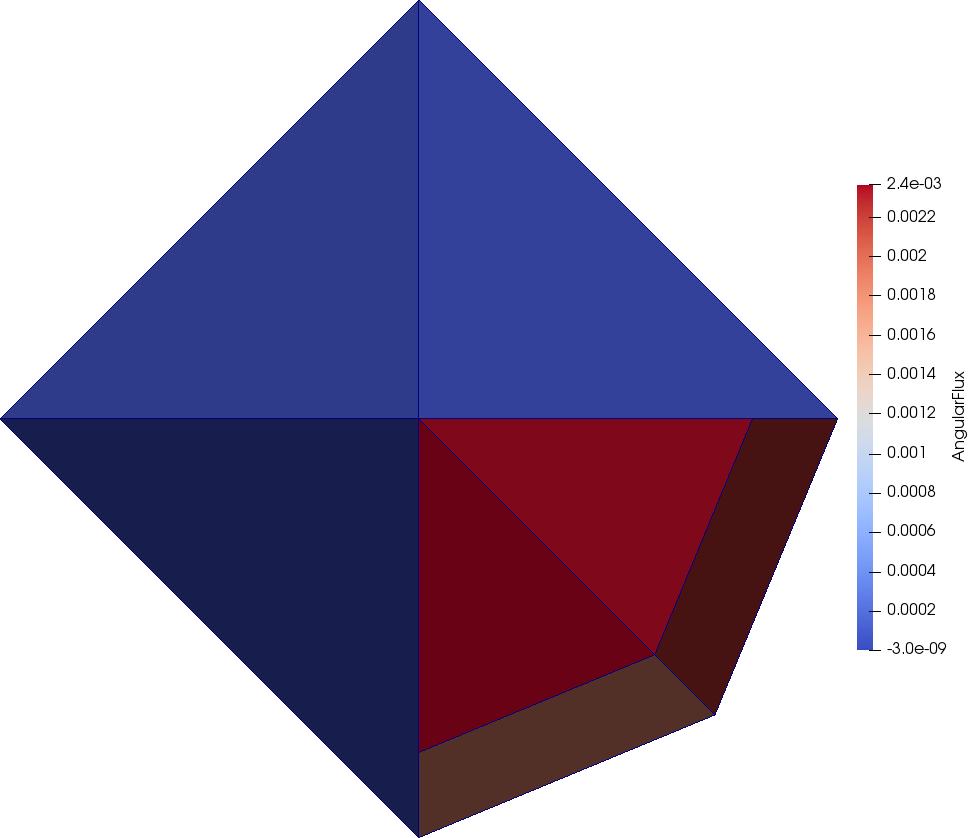}}
\caption{Angular flux at four spatial points after 2 regular adapt steps in a 2D pure streaming problem with a small source at the centre of the domain. The four spatial points correspond to the green dots shown in \fref{fig:no_angles_2}. The P$^0$ angular discretisation is on the $r=1$ sphere, but has been projected onto faceted polyhedra for ease of visualisation. The camera is pointed in the $-z$ direction.}
\label{fig:ang_flux_adapt_2}
\end{figure}

\begin{table}[ht]
\centering
\begin{tabular}{ c c c | c c c c c c c}
\toprule
CG nodes & Adapt step. & NDOFs & $n_\textrm{its}$ & CC & Op Complx &  WUs$^\textrm{full}$ & WUs$^\textrm{DG}$ & Memory\\
\midrule
2313 & 1 & 6.3\xten{4} & 17 & 5.1 & 2.7 & 115 & 27.3 & 10.9 \\
2313 & 2 & 9.2\xten{4} & 16 & 5.3 & 2.8 & 112 & 26.9 & 11.3 \\
2313 & 3 & 2.2\xten{5} & 17 & 5.8 & 3.0 & 128 & 30.5 & 11.9 \\
2313 & 4 & 2.8\xten{5} & 16 & 5.2 & 2.8 & 111 & 26.5 & 11.2 \\
2313 & 5 & 3.1\xten{5} & 19 & 5.3 & 2.8 & 130 & 31.0 & 11.2 \\
\bottomrule  
\end{tabular}
\caption{Results from using AIRG on a pure streaming problem in 2D with CF splitting by directional element agglomeration, drop tolerance on $\mat{A}$ of 0.0075 and $\mat{R}$ of 0.025, with regular angular adaptivity with a refinement tolerance of 0.001, a maximum of 3 levels of angular refinement and 5 adapt steps. The WUs listed are scaled by the nnzs in the \textit{adapted} solve at each step.}
\label{tab:2D_stream_adapt_element}
\end{table}
\tref{tab:2D_stream_adapt_element} shows the results from using a regular angle adapt with fixed spatial resolution (this is the third refined mesh from \cite{Dargaville2023}) in the streaming limit, with directional element agglomeration to compute the CF splitting. \fref{fig:directional_coarsening} shows the scalar flux and the distribution of our angular resolution throughout space. Our adapt process is focusing resolution in the directions away from the source in order to resolve the rays-effects in this problem. Using a regular adapt process in this problem is not necessary, given the optimal angular resolution can be easily determined \textit{a priori} (refining in directions looking away from the source), but we wished to show the effect of solving our adapted discretisations at different adapt steps. \fref{fig:directional_coarsening} also shows that after the first adapt step, the directional element agglomeration is gluing together elements in the ``important'' direction at each spatial node, given by the refined angular flux in \fref{fig:ang_flux_adapt_2}. We see this results in an iteration count and overall work that is close to flat, with approximately 15\% growth after 5 adapt steps. Interestingly, if we force the element agglomeration to be directionless at every adapt step, we see very similar cycle and operator complexities, with 15, 16, 17, 17 and 18 iterations. This implies that the directional element agglomeration is unnecessary. Further investigation revealed that it is the high GMRES polynomial order in AIRG ($m=4$, i.e., a third order polynomial) that is compensating for the directionless CF splitting; reducing the polynomial order to 1 with the directional CF splitting gives 17, 31, 44, 53, 65 iterations compared with directionless at 17, 34, 40, 56 and 81 iterations. Our strong approximations to the ideal operators and F-point smoothers are compensating for the poorer CF splitting. 

With higher spatial resolution (not pictured), the adapt can also remove the angular resolution in the regions between the rays that are added in the initial adapt steps; the coarse spatial resolution used in \fref{fig:directional_coarsening} means that numerical diffusion keeps some of the angular resolution above the removal threshold. Even with this excess resolution, we have 2.8$\times$ and 4.8 fewer NDOFs for the adapted solves in step 2 and 3 than in a uniform level 2 and 3 discretisation in this problem. Similarly, there are 2.75 and 4.6$\times$ fewer nnzs in the adapted matrices compared to the uniform. This shows that the pathology described in \secref{sec:sub-grid} where the nnzs grows considerably with the adapt is not seen in this problem and that the sparsity control in \secref{sec:iterative} is effective; the NDOFs grow 4.74$\times$ from adapt step 1 to 5, with the nnzs growing 4.88$\times$, or an increase of around 3\% above the NDOFs. 

\begin{table}[ht]
\centering
\begin{tabular}{ c c c | c c c c c c c}
\toprule
CG nodes & Adapt step. & NDOFs & $n_\textrm{its}$ & CC & Op Complx &  WUs$^\textrm{full}$ & WUs$^\textrm{DG}$ & Memory\\
\midrule
2313 & 1 & 6.3\xten{4} & 9 & 4.4 & 2.87 & 60 & 14.1 & 10.2 \\
2313 & 2 & 9.2\xten{4} & 11 & 4.7 & 3.2 & 74 & 17.6 & 10.7 \\
2313 & 3 & 2.2\xten{5} & 12 & 5.0 & 3.6 & 85 & 20.2 & 11.3 \\
2313 & 4 & 2.8\xten{5} & 11 & 4.6 & 3.3 & 73 & 17.5 & 10.7 \\
2313 & 5 & 3.1\xten{5} & 11 & 4.5 & 3.2 & 72 & 17.2 & 10.6 \\
\bottomrule  
\end{tabular}
\caption{Results from using AIRG on a pure streaming problem in 2D with CF splitting by the \textit{hypre} implementation of Falgout-CLJP with a strong threshold of 0.2, drop tolerance on $\mat{A}$ of 0.0075 and $\mat{R}$ of 0.025, with regular angular adaptivity with a refinement tolerance of 0.001, a maximum of 3 levels of angular refinement and 5 adapt steps. The WUs listed are scaled by the nnzs in the \textit{adapted} solve at each step.}
\label{tab:2D_stream_adapt}
\end{table}
\tref{tab:2D_stream_adapt} shows the results from using the Falgout-CLJP CF splitting instead of our directionless agglomeration algorithm, and we see much lower iteration count with similarly flat work at each adapt step, with fixed memory use of 10-11 copies of the angular flux.. This shows that accounting for the directionality of all the angles (not just the ``important'' angles) can improve the convergence in the streaming limit. Both the directional element agglomeration and Falgout-CLJP CF splitting however result in close to fixed work in this streaming problem when adapting. This shows the power of combining our iterative method and P$^0$ angular adaptivity and is in contrast to the results in \cite{Dargaville2019}, where we saw that after 3 angular adapt steps in a streaming problem, solving in Haar space with a matrix-free multigrid resulted in an iteration count that grew from 25 to 91.

\begin{figure}[ht]
\centering
\includegraphics[width =0.45\textwidth]{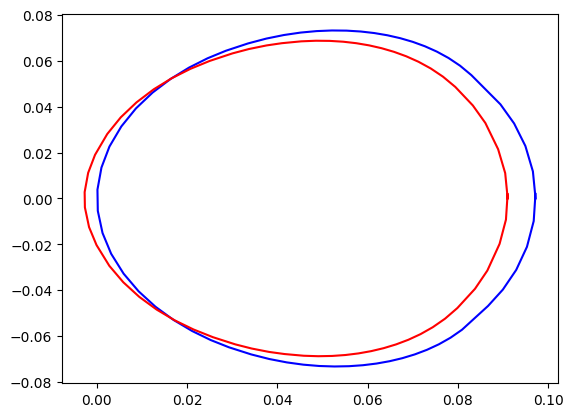}
\caption{Field of values of the streaming operators on the third spatial grid. Blue is uniform level 3 angular refinement, red is an adapted angular discretisation with a maximum level of refinement of 3.}
\label{fig:streaming_adapt}
\end{figure}

In general with our adapted matrices, we might expect similar spectrums when compared to the uniform matrices (i.e., when comparing a uniform discretisation to that of an adapt with the same maximum level of refinement) and therefore require a similar (or smaller) number of iterations to converge. Typically this is the case. \tref{tab:2D_stream_adapt} however shows that with Falgout-CLJP CF splitting, the iteration count starts at 9 given the adapt process starts with a uniform level 1 solve, the second and third steps however takes 11 and 12 iterations. The results in \cite{Dargaville2023} show that with uniform level 2 and 3 discretisations this problem took 10 and 11 iterations to solve, respectively. The uniform level 3 matrix on this mesh has a condition number of approx. 560 compared with around 1780 for the adapted discretisation in step 3 pictured in \fref{fig:solution_adapt_3}, resulting in the increase in iteration count. \fref{fig:streaming_adapt} shows that the field of values for the adapted discretisation is closer to the origin when compared to the uniform, and the convergence of our GMRES polynomials is affected by this \cite{Meurant2020}. As noted the adapted discretisation has fewer DOFs (and nnzs) so it is still much cheaper to solve than the uniform discretisation, but it is worth noting that solving the matrices generated by our adapt steps is not always equivalent to solving a uniform discretisation. 

\subsubsection{Scattering problem}
\label{sec:pure_scatter_adapt}
\begin{table}[ht]
\centering
\begin{tabular}{ c c c | c c c c c c c}
\toprule
CG nodes & Adapt step. & NDOFs & $n_\textrm{its}$ & CC & Op Complx &  WUs$^\textrm{mf}$ & WUs$^\textrm{DG}$ & Memory\\
\midrule
2313 & 1 & 6.3\xten{4} & 25 & 3.9 & 2.0 & 38 & 69 & 16.9 \\
2313 & 2 & 2.4\xten{5} & 27 & 3.8 & 2.0 & 39 & 72 & 16.5 \\
2313 & 3 & 3.1\xten{5} & 27 & 3.8 & 2.0 & 38 & 76 & 16.5 \\
\bottomrule  
\end{tabular}
\caption{Results from using additive preconditioning on a pure scattering problem with total and scattering cross-section of 10.0 in 2D with regular angular adaptivity with a refinement tolerance of 0.001, a maximum of 3 levels of angular refinement and 3 adapt steps. The WUs listed are scaled by the nnzs in the \textit{adapted} solve at each step. The cycle and operator complexity listed are for AIRG on $\mat{M}_\Omega$ with CF splitting by element agglomeration.}
\label{tab:2D_stream_adapt_diffusion_element}
\end{table}

\begin{figure}[h]
\centering
\subfloat[][Scalar flux]{\label{fig:solution_adapt_1_diffusion}\includegraphics[width =0.25\textwidth]{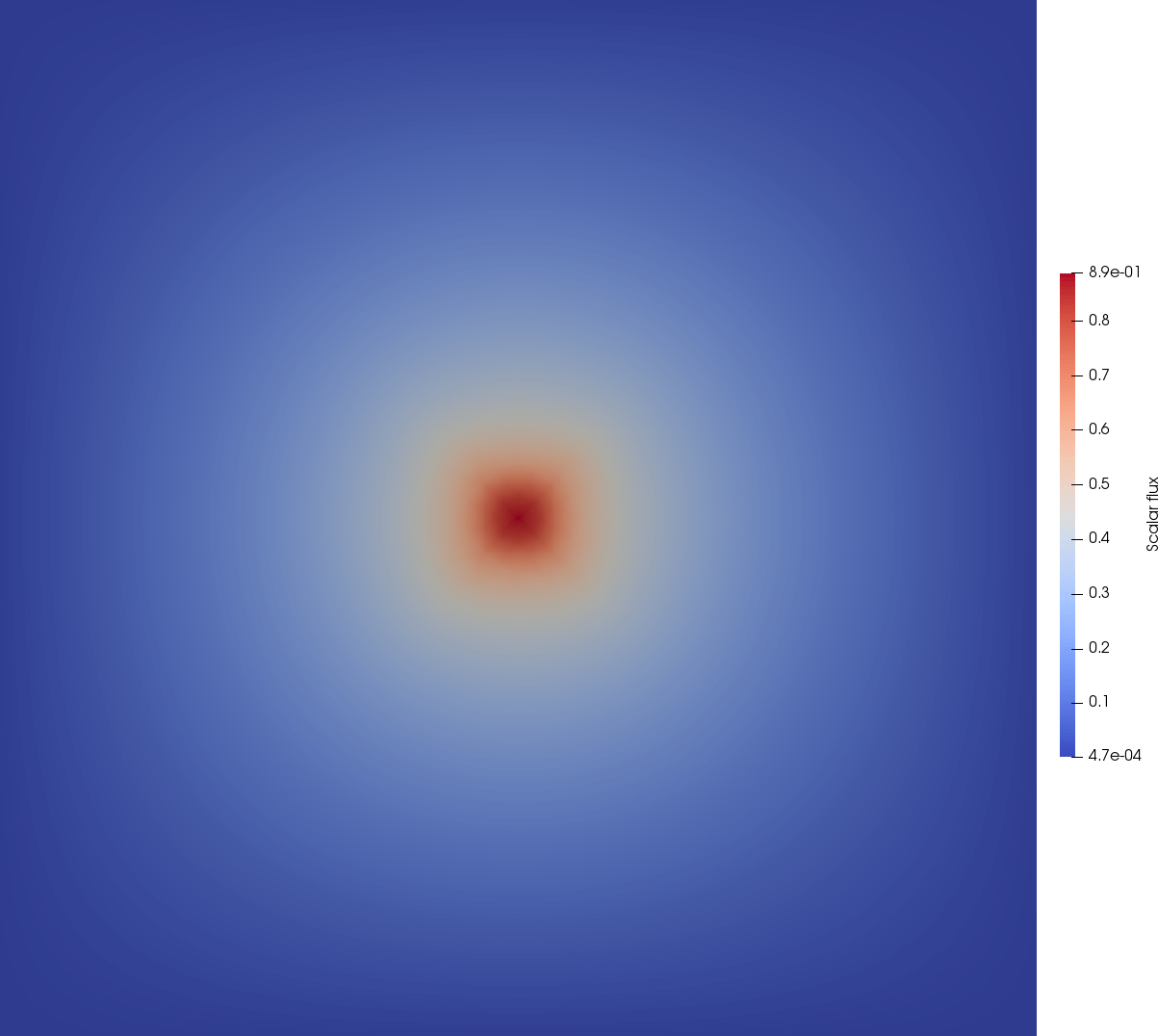}} \quad
\subfloat[][Number of angles across space]{\label{fig:no_angles_1_diffusion}\includegraphics[width =0.25\textwidth]{no_angles_1.png}}
\subfloat[][Element agglomeration on the second multigrid level]{\label{fig:coarsening_1_diffusion} \quad\includegraphics[width =0.225\textwidth]{coarsening_1.png}}\\
\subfloat[][]{\label{fig:solution_adapt_2_diffusion}\includegraphics[width =0.25\textwidth]{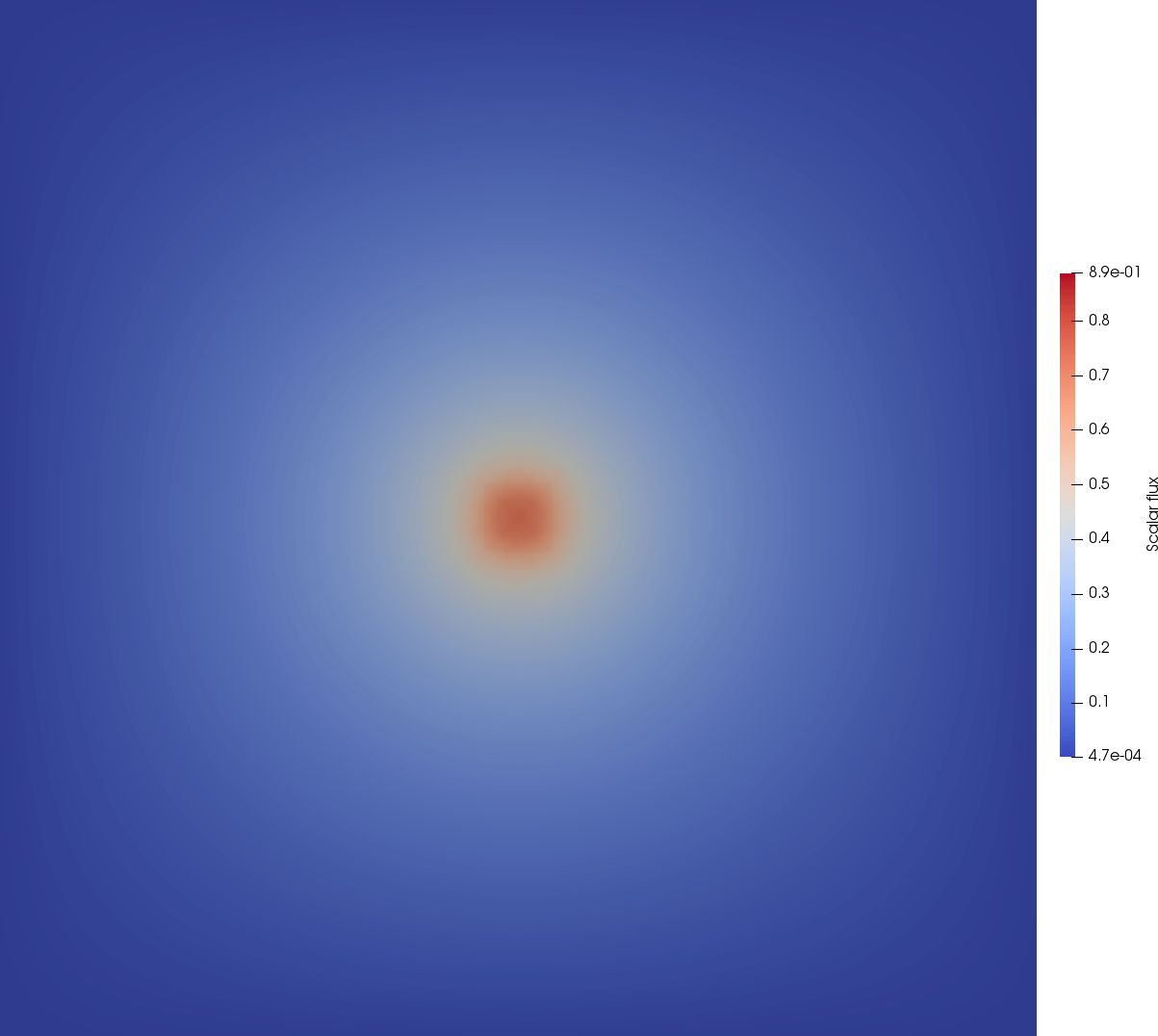}} \quad
\subfloat[][]{\label{fig:no_angles_2_diffusion}\includegraphics[width =0.25\textwidth]{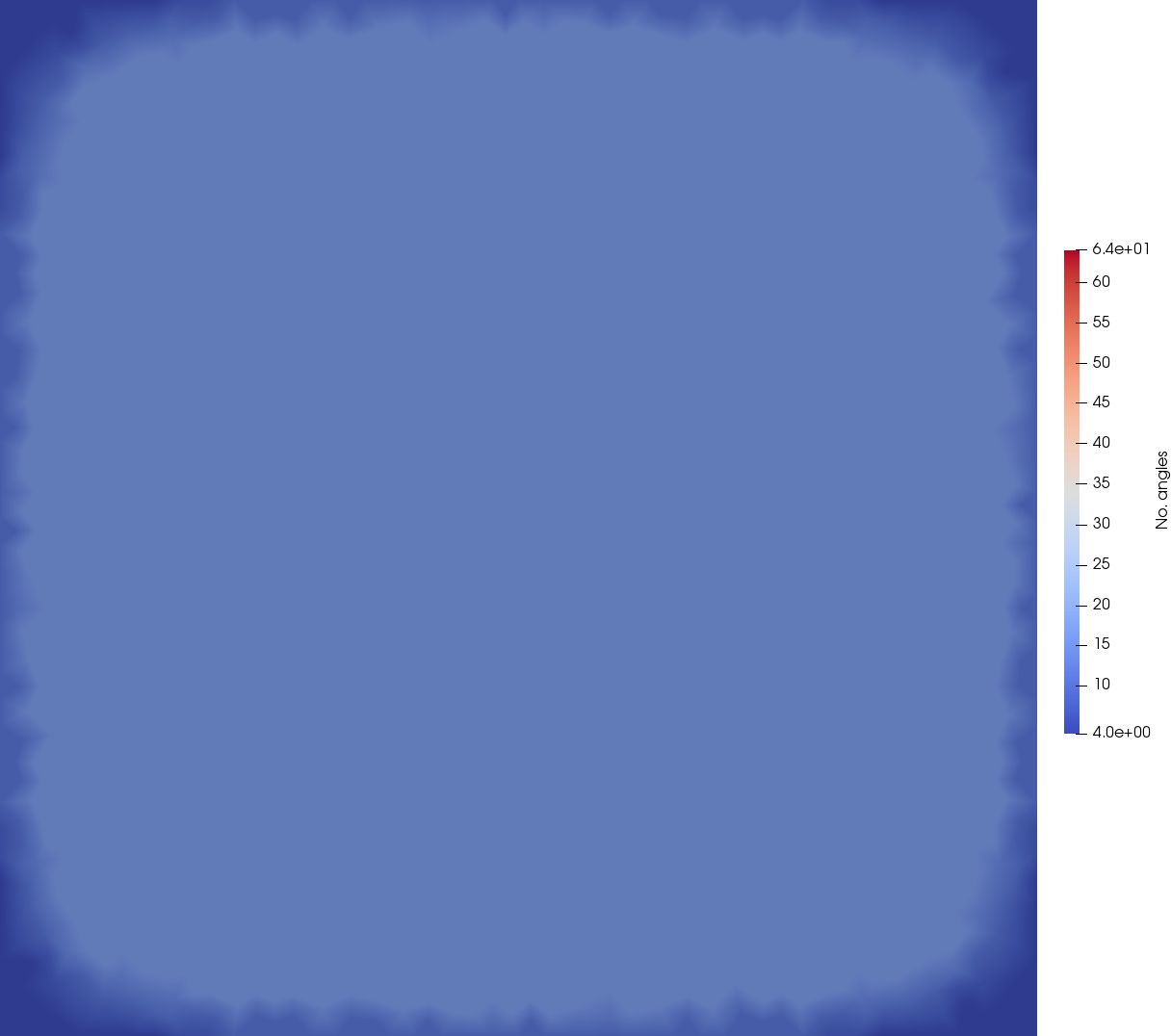}}
\subfloat[][]{\label{fig:coarsening_2_diffusion} \quad\includegraphics[width =0.225\textwidth]{coarsening_1.png}}\\
\subfloat[][]{\label{fig:solution_adapt_3_diffusion}\includegraphics[width =0.25\textwidth]{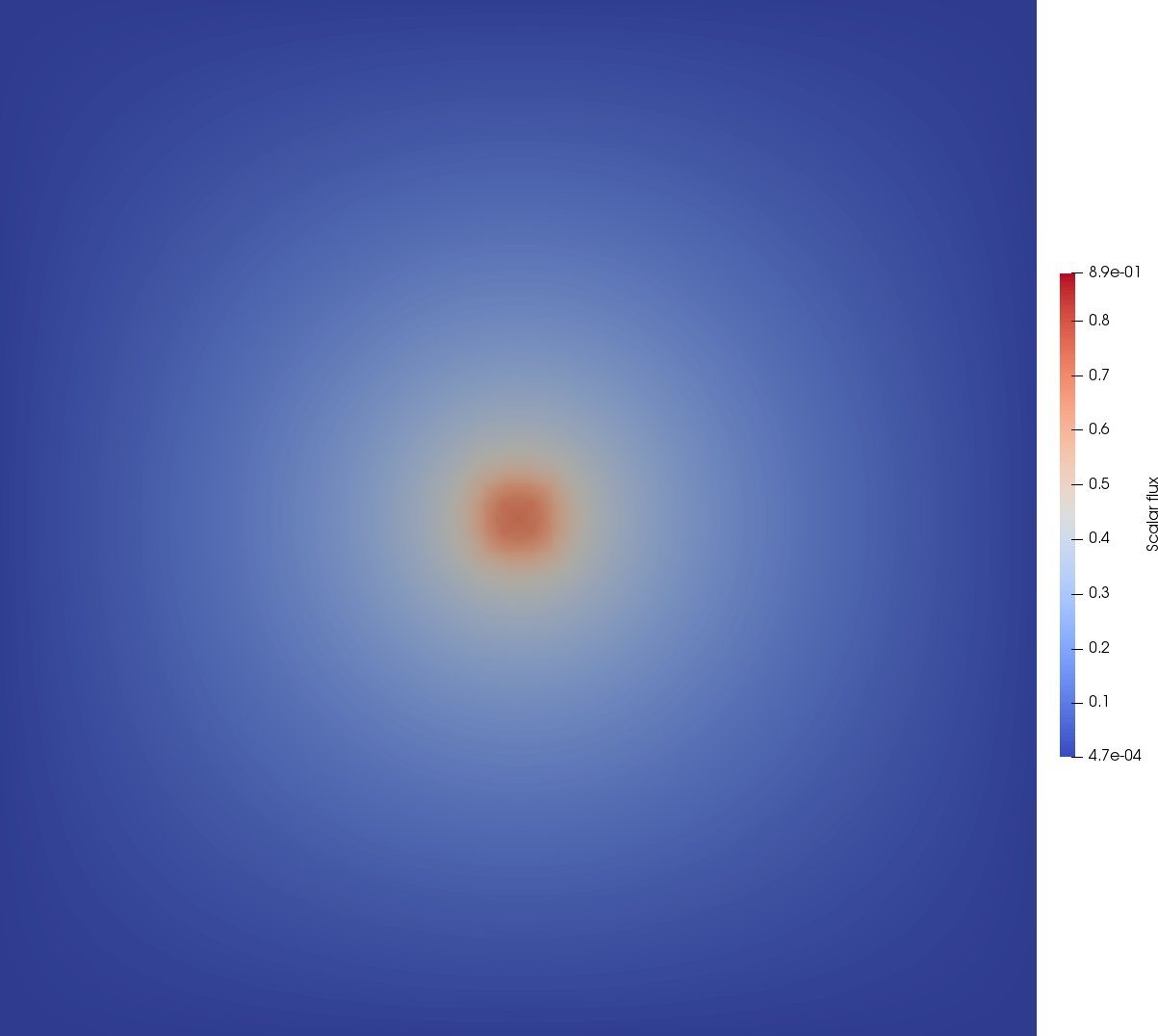}} \quad
\subfloat[][The green dots correspond to the angular flux output in \fref{fig:ang_flux_adapt_diffusion_3}]{\label{fig:no_angles_3_diffusion}\includegraphics[width =0.25\textwidth]{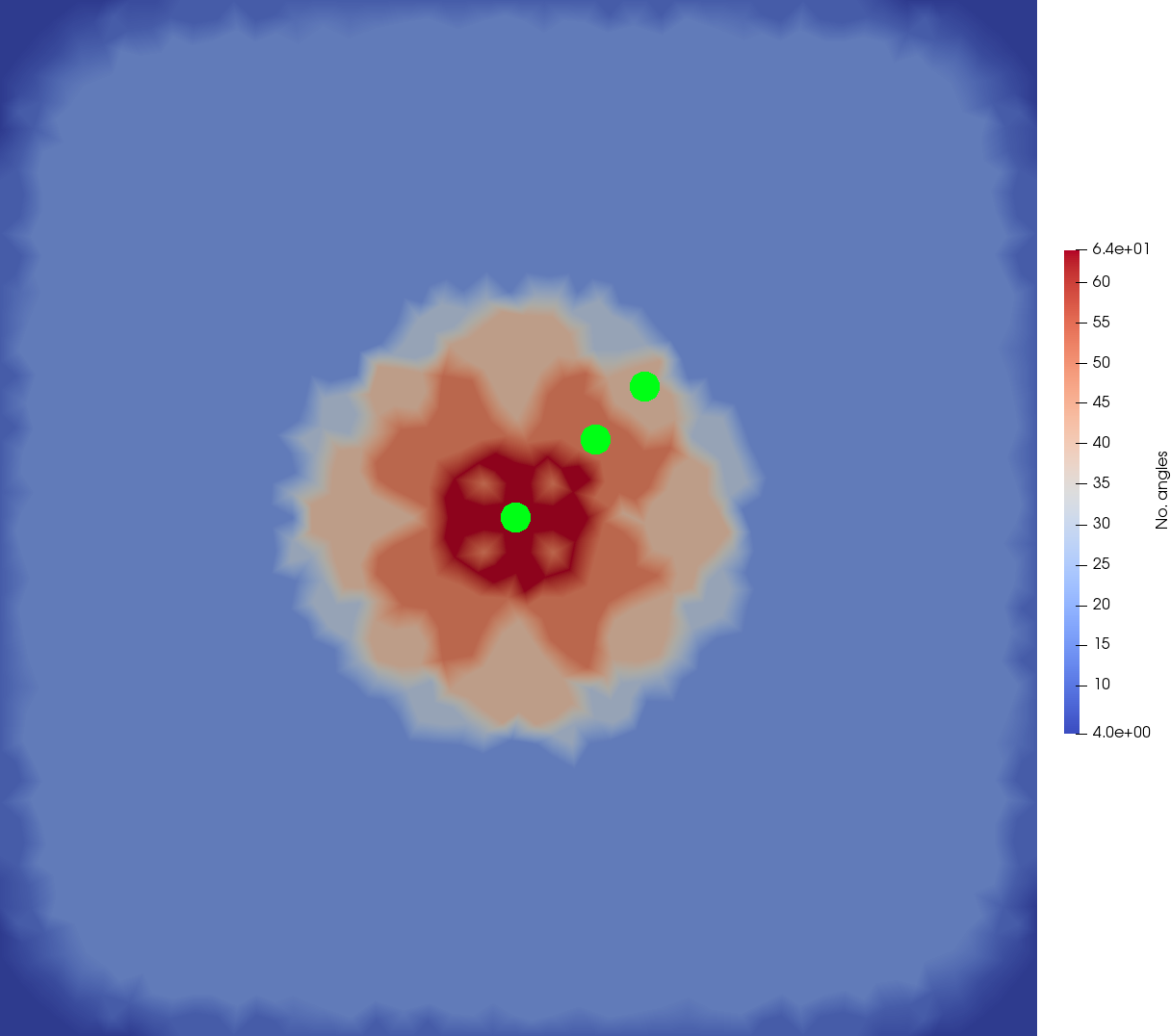}}
\subfloat[][]{\label{fig:coarsening_3_diffusion} \quad\includegraphics[width =0.225\textwidth]{coarsening_1.png}}\\
\caption{Adapt results for a 2D pure scattering problem with total and scatter cross-section of 10.0 with a small source at the centre of the domain, allowing a maximum of 3 levels of regular angular refinement (giving a max. number of angles as 64). The first column shows the scalar flux and the second the number of angles across space. The third column shows the resulting element agglomeration on the second spatial grid if the directional algorithm in \secref{sec:Element agglomeration methods} is used. The rows show consecutive regular adapt steps}
\label{fig:directional_coarsening_diffusion}
\end{figure}

We now examine the performance of our additively preconditioned iterative method with adaptivity in the scattering limit. We only require three adapt steps to resolve the required resolution in this problem; \fref{fig:directional_coarsening_diffusion} shows this adapt process and we can see the majority of the angular resolution in the problem is focused around the source. \fref{fig:directional_coarsening_diffusion} also shows that given the high scattering cross-section, the element agglomeration has proceeded uniformly, with no directional information after the first adapt step. \tref{tab:2D_stream_adapt_diffusion_element} shows that with the CF splitting by element agglomeration, we see a plateau in the iteration count and work. In contrast to the pure streaming problem in \secref{sec:pure_stream_adapt}, performing the CF splitting with Falgout-CLJP results in an identical iteration count (see \tref{tab:2D_stream_adapt_diffusion}), with slightly higher work due to the higher cycle complexity. This confirms that although the element agglomeration results in poorer CF splittings, the streaming/removal operator is easier to invert and in problems with a large removal term, a simple CF splitting algorithm without access to the matrix entries can be sufficient to ensure scalability. 

For both methods, the iteration count for each of the adapt steps is lower or equal to that of the uniform angular refinement shown in \cite{Dargaville2023}, with the work and memory use constant at around 38 WUs and 17 copies of the angular flux, respectively. The NDOFs grows 4.96$\times$ from adapt step 1 to 3, with the growth in the nnzs of the streaming/removal operator at 5.07, giving an increase of around 2\%.
\begin{table}[ht]
\centering
\begin{tabular}{ c c c | c c c c c c c}
\toprule
CG nodes & Adapt step. & NDOFs & $n_\textrm{its}$ & CC & Op Complx &  WUs$^\textrm{mf}$ & WUs$^\textrm{DG}$ & Memory\\
\midrule
2313 & 1 & 6.3\xten{4} & 25 & 4.1 & 1.7 & 38 & 69 & 17.2 \\
2313 & 2 & 2.4\xten{5} & 27 & 4.0 & 1.4 & 39 & 73 & 16.5 \\
2313 & 3 & 3.1\xten{5} & 27 & 4.1 & 1.4 & 38 & 77 & 16.5 \\
\bottomrule  
\end{tabular}
\caption{Results from using additive preconditioning on a pure scattering problem with total and scattering cross-section of 10.0 in 2D with regular angular adaptivity with a refinement tolerance of 0.001, a maximum of 3 levels of angular refinement and 3 adapt steps. The WUs listed are scaled by the nnzs in the \textit{adapted} solve at each step. The cycle and operator complexity listed are for AIRG on $\mat{M}_\Omega$ with CF splitting by Falgout-CLJP.}
\label{tab:2D_stream_adapt_diffusion}
\end{table}

\begin{figure}[ht]
\centering
\subfloat[][$x=1.5, y=1.5$]{\label{fig:middle_3}\includegraphics[width =0.25\textwidth]{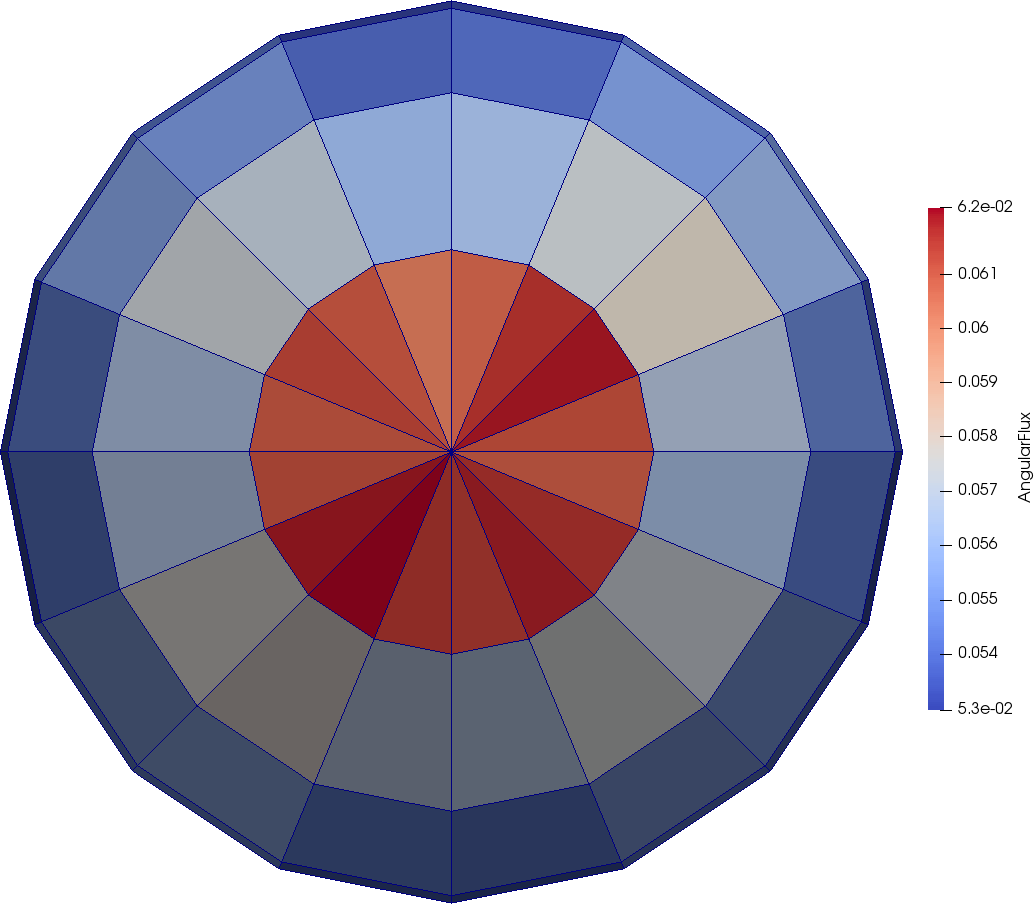}} \quad
\subfloat[][$x=1.72963, y=1.76576$]{\label{fig:further_out_3}\includegraphics[width =0.25\textwidth]{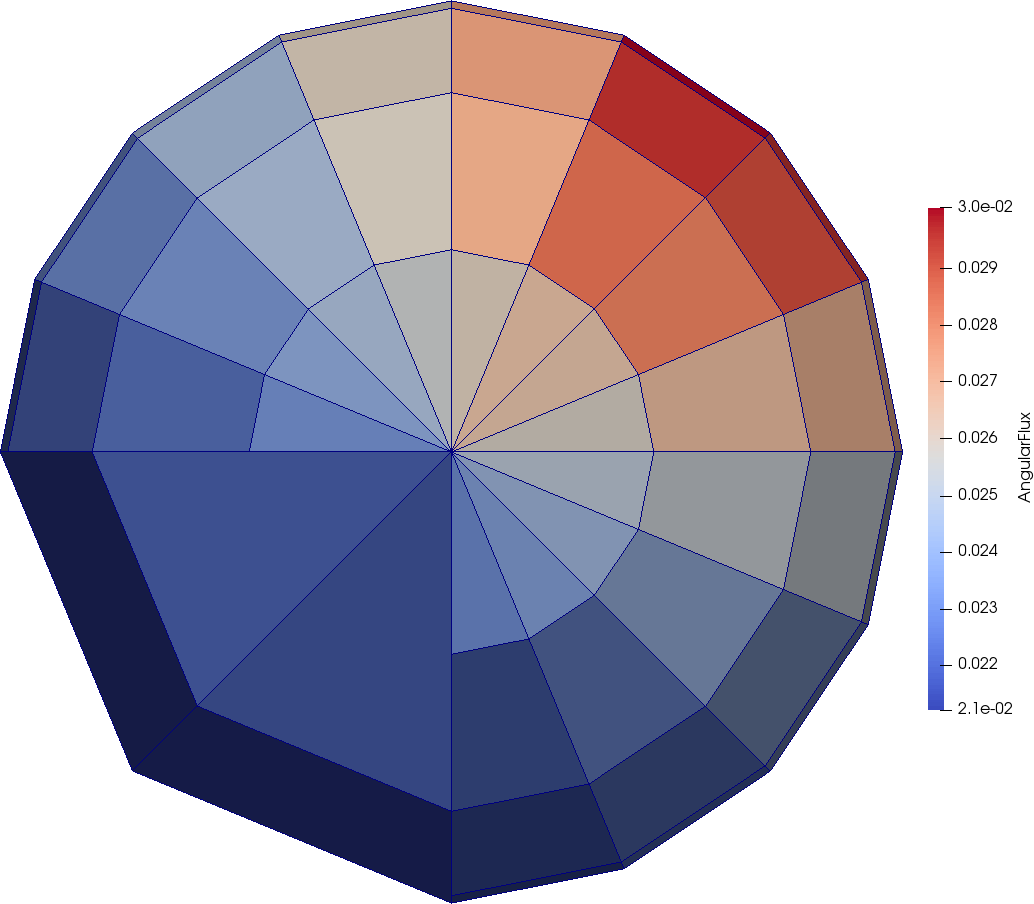}}
\subfloat[][$x=1.83372, y=1.87562$]{\label{fig:furthest_out_3} \quad\includegraphics[width =0.25\textwidth]{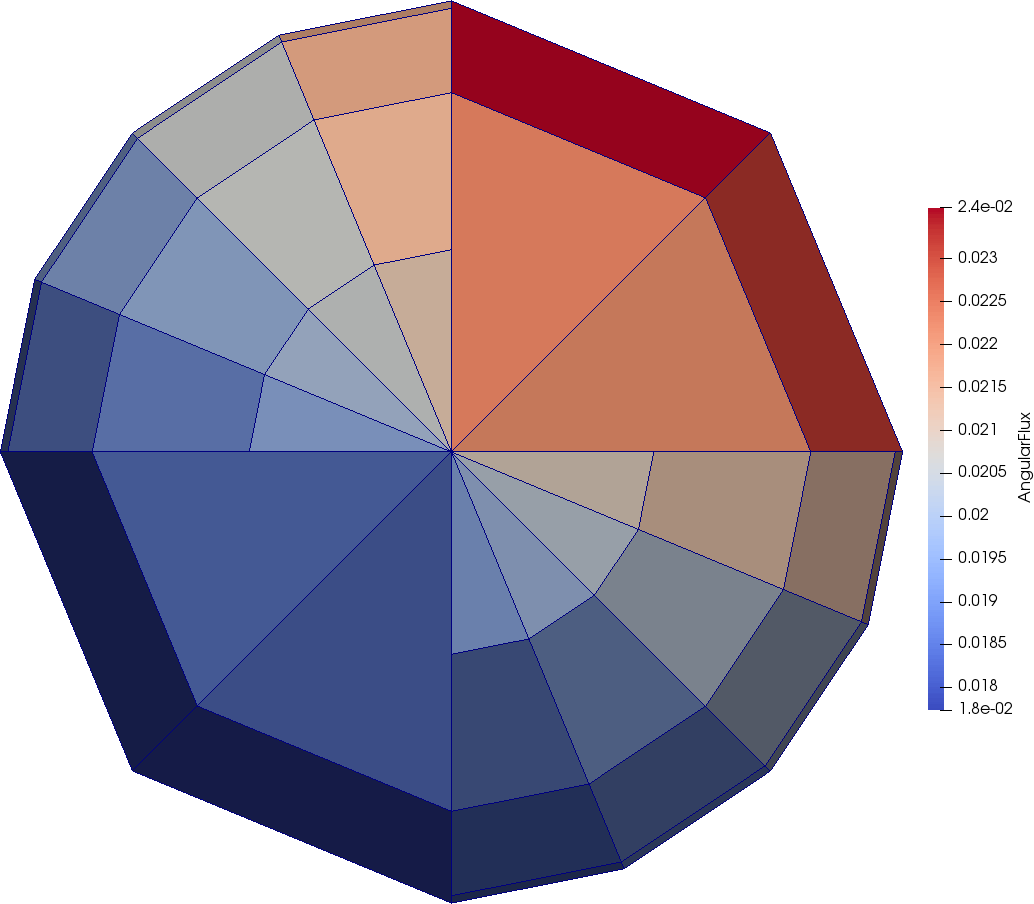}}
\caption{Angular flux at three spatial points after 3 regular adapt steps in a 2D pure scattering problem with a small source at the centre of the domain. The three spatial points correspond to the green dots shown in \fref{fig:no_angles_3_diffusion}. The P$^0$ angular discretisation is on the $r=1$ sphere, but has been projected onto faceted polyhedra for ease of visualisation. The camera is pointed in the $-z$ direction.}
\label{fig:ang_flux_adapt_diffusion_3}
\end{figure}
\subsubsection{Reduced tolerance adapts}
\label{sec:reduced_tol}
In the previous sections we have shown that AIRG and our additively preconditioned iterative method can solve adapted P$^0$ problems in both the streaming and scattering limit with close to fixed work with a zero initial condition used at each adapt step. In practice however we would like to use the solution from the previous adapt step as an initial condition to reduce the number of iterations; in streaming problems where the ray-effects between different adapt steps do not align we find this doesn't change the convergence. In scattering problems however it can help reduce the iteration count.

Furthermore, our previous work has used reduced tolerance solves \cite{Dargaville2019, Dargaville2019a, Dargaville2019b} to decrease the cost of our adaptive process, where the linear system at each adapt step, except the last, is solved to a bigger tolerance, typically 1\xten{-3} in the 2-norm; this is not very robust and requires problem specific tuning. We noted in \cite{Dargaville2019} however that we only need to converge sufficiently such that the error in each of our wavelets (this is also true for any hierarchical angular discretisation, such as P$_n$ or FP$_n$) has converged to within a relative tolerance of 0.1, so that we can determine whether the error is greater than the refinement threshold of 1.0 defined in \secref{sec:ang_discs} (and equivalent for de-refinement with a threshold of 0.01). The convergence criteria for the iterative method during each adapt step, except the last, can therefore be set as the infinity norm on the relative error being less than 0.1. 

With goal-based adaptivity this requires that the error metric used has a very good effectivity index and we are forced to compute the goal-based error metric at each iteration which may be expensive. For regular adaptivity however this is very simple, as the error is given by a scaled version of our P$^0$ solution mapped to wavelet space. As such, with our P$^0$ discretisation we need to map to the equivalent wavelet space (at the cost of an $\mathcal{O}(n)$ mapping) and compute the relative change of each wavelet coefficient; if all the wavelet coefficients have converged to a relative tolerance of 0.1 our iterative method has converged sufficiently for this adapt step and we can exactly determine the refinement/de-refinement required by the next adapt step. 
\begin{figure}[ht]
\centering
\subfloat[][Scalar flux]{\label{fig:scalar_flux_threshold}\includegraphics[width =0.35\textwidth]{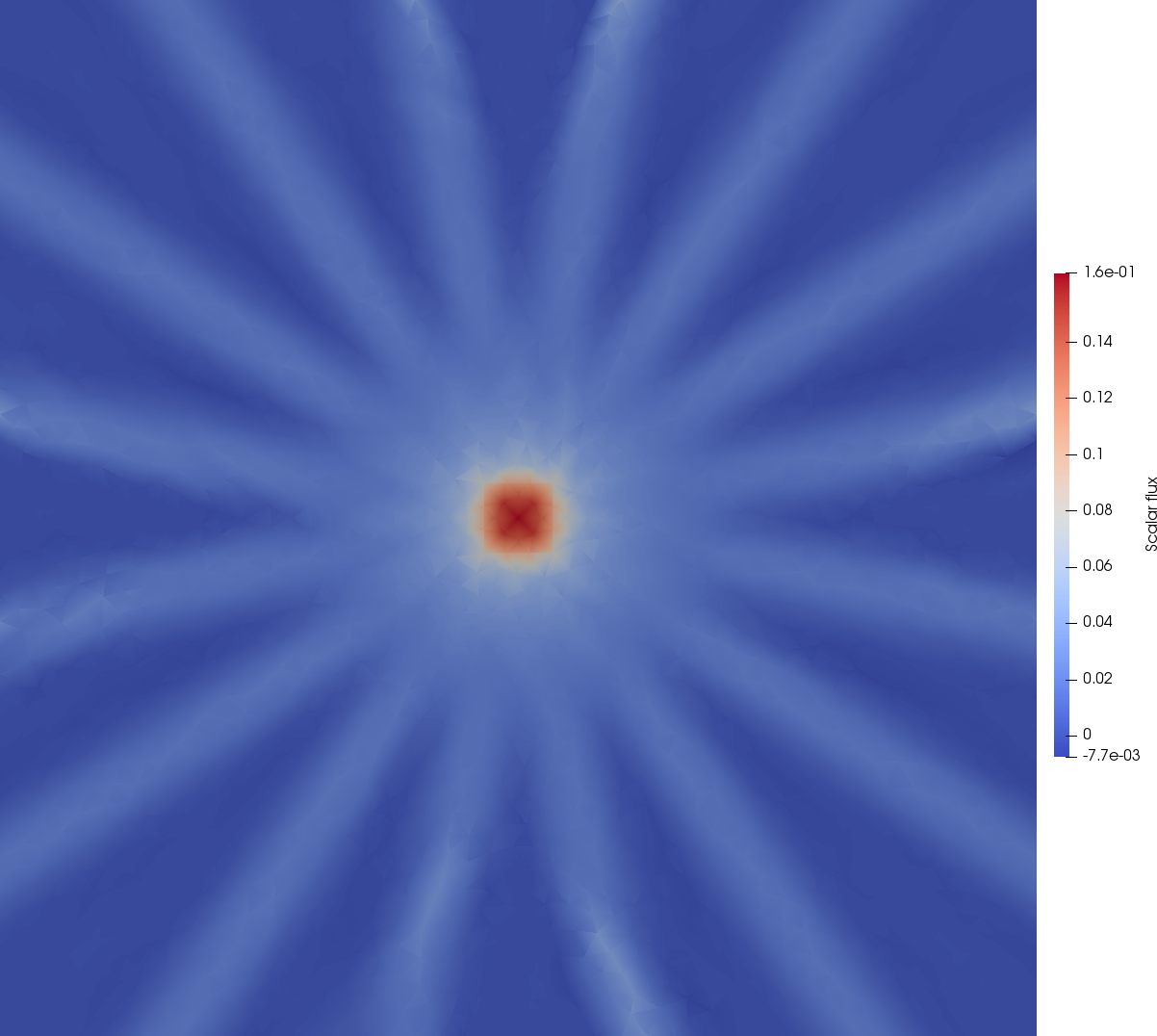}} \quad
\subfloat[][Number of angles across space]{\label{fig:no_angles_threshold}\includegraphics[width =0.35\textwidth]{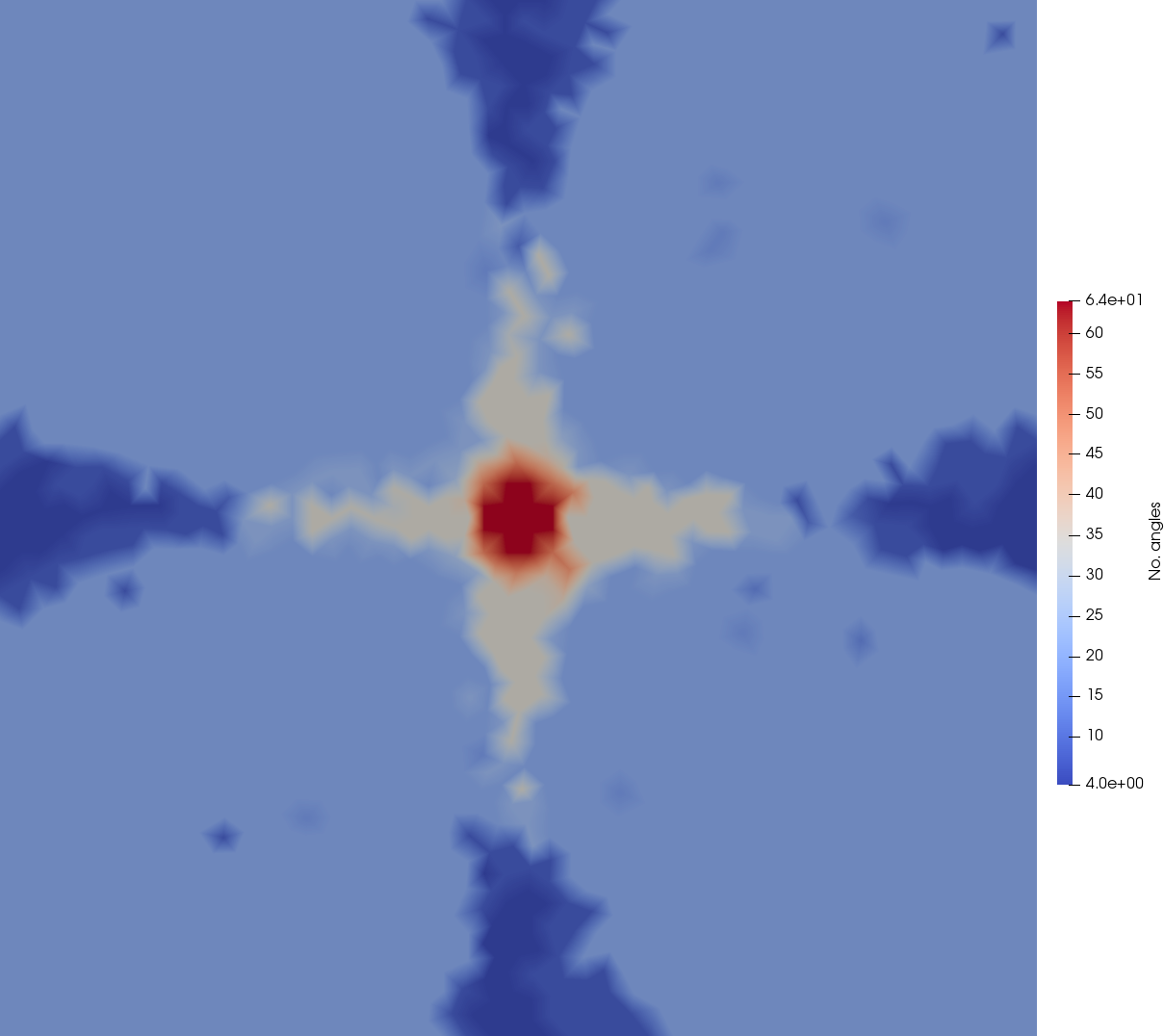}}\\
\caption{Adapt results after 5 regular adapt steps in a 2D pure streaming problem with a small source at the centre of the domain, allowing a maximum of 3 levels of regular angular refinement, with adapt steps prior to the final step solved with wavelet-based reduced tolerance to determine convergence.}
\label{fig:threshold_stream}
\end{figure}

This is made more difficult by our sub-grid scale discretisation, as the error calculation and thresholding after each adapt step uses $\bm{\Psi}$, and hence requires computing $\bm{\Theta}$ with \eref{eq:theta} at every iteration which is costly given the matrix multiplications required. Instead, we map the ``coarse'' variable $\bm{\Phi}$ to wavelet space at each iteration and check the convergence of the wavelet coefficients on the continuous mesh. This doesn't necessarily produce the exact adapted discretisation that would otherwise have been built, but it is very close; there are typically only a handful of wavelets that would otherwise be refined/de-refined at any single adapt step and those can be picked up by subsequent steps. \fref{fig:threshold_stream} shows the results from using this reduced tolerance in a streaming problem and we can see that after 5 adapt steps, this results in a scalar flux and adapted discretisation almost identical to that without the reduced tolerance solve, shown in Figures \ref{fig:solution_adapt_5} and \ref{fig:no_angles_4}. We can see in \fref{fig:threshold_stream} that there is a small region at around $x=1.5, y=0$ where refinement has not been triggered in the same way, but overall we find this is a very robust way to reduce the cost of our adaptive process. 

\tref{tab:2D_stream_adapt_reduced} shows the results from using both a non-zero intial guess (which has very little effect) and the reduced tolerance solves in a streaming problem. We should note we have included the extra cost that comes from mapping to wavelet space and computing the relative change in each wavelet coefficient at every iteration in the WUs. Compared to \tref{tab:2D_stream_adapt} we can see the cost of the first four adapt steps has been reduced considerably. If we compare the total cost of all five adapt steps and scale by the nnzs in a uniform level 3 discretisation, we find the total work has reduced from 72 WUs to 54. \cite{Dargaville2023} showed that it costs 70 WUs to solve a uniform level 3 discretisation, so our adapt process with a reduced tolerance solve beats the uniform equivalent. This cost saving only increases as the number of adapt steps is increased; previously in \cite{Dargaville2019, Dargaville2019a, Dargaville2019b} we typically needed a higher number of adapt steps than five in order to beat a uniform discretisation. 

\begin{table}[ht]
\centering
\begin{tabular}{ c c c | c c c c c c c}
\toprule
CG nodes & Adapt step. & NDOFs & $n_\textrm{its}$ & CC & Op Complx &  WUs$^\textrm{full}$ & WUs$^\textrm{DG}$ & Memory\\
\midrule
2313 & 1 & 6.3\xten{4} & 2 & 4.4 & 2.87 & 22 & 5.3 & 10.2 \\
2313 & 2 & 9.2\xten{4} & 6 & 4.7 & 3.3 & 47 & 11.3 & 10.6 \\
2313 & 3 & 2.2\xten{5} & 7 & 5.0 & 3.6 & 56 & 13.4 & 11.1 \\
2313 & 4 & 2.8\xten{5} & 7 & 4.6 & 3.3 & 53 & 12.6 & 10.6 \\
2313 & 5 & 3.1\xten{5} & 10 & 4.5 & 3.2 & 66 & 15.9 & 10.4 \\
\bottomrule  
\end{tabular}
\caption{Results from using AIRG on a pure streaming problem in 2D with CF splitting by the \textit{hypre} implementation of Falgout-CLJP with a strong threshold of 0.2, drop tolerance on $\mat{A}$ of 0.0075 and $\mat{R}$ of 0.025, with regular angular adaptivity with a refinement tolerance of 0.001, a maximum of 3 levels of angular refinement, 5 adapt steps and with adapt steps prior to the final step solved with wavelet-based reduced tolerance to determine convergence. The WUs listed are scaled by the nnzs in the \textit{adapted} solve at each step.}
\label{tab:2D_stream_adapt_reduced}
\end{table}

\begin{figure}[ht]
\centering
\subfloat[][Scalar flux]{\label{fig:scalar_flux_threshold_diffusion}\includegraphics[width =0.35\textwidth]{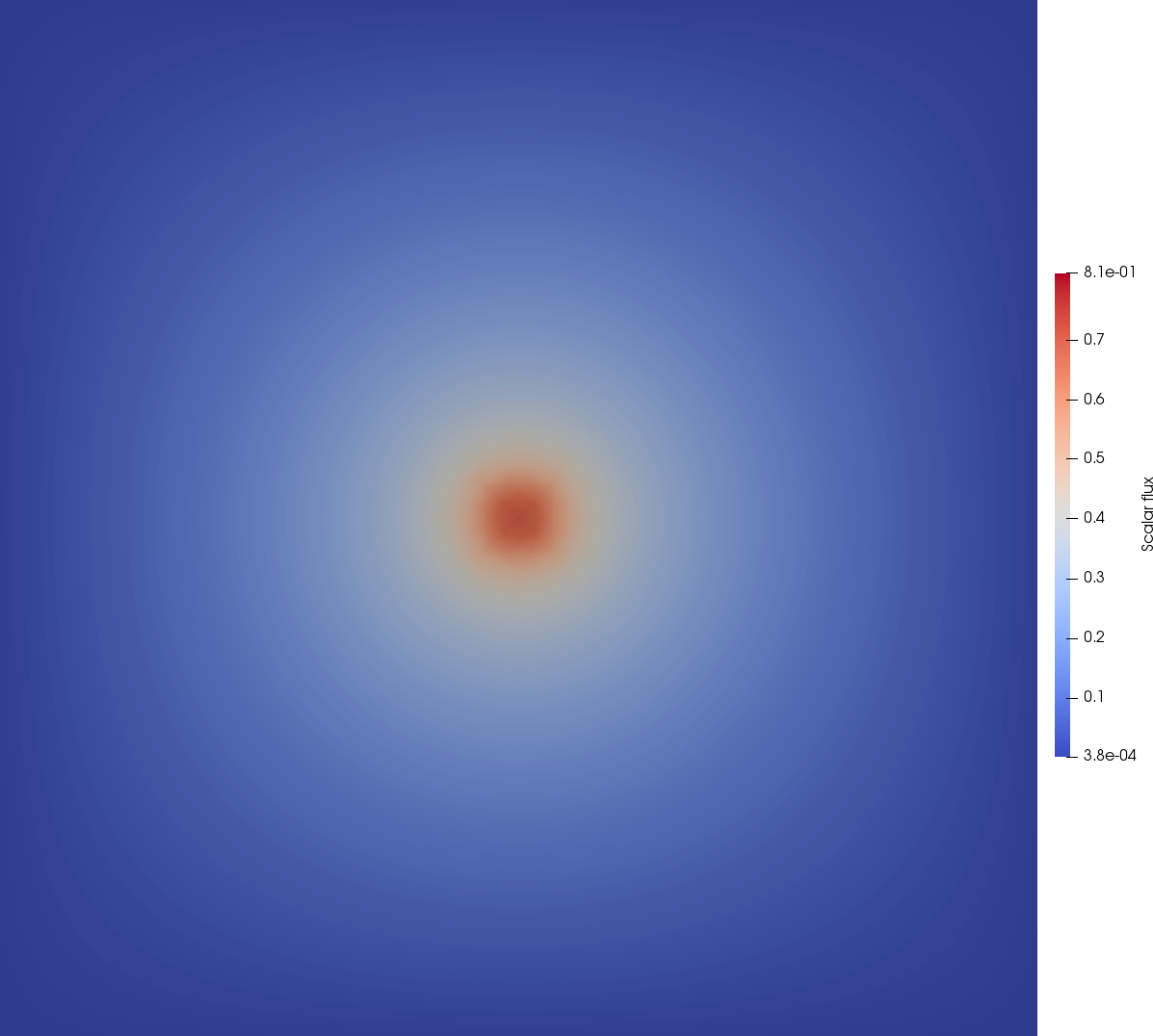}} \quad
\subfloat[][Number of angles across space]{\label{fig:no_angles_threshold_diffusion}\includegraphics[width =0.35\textwidth]{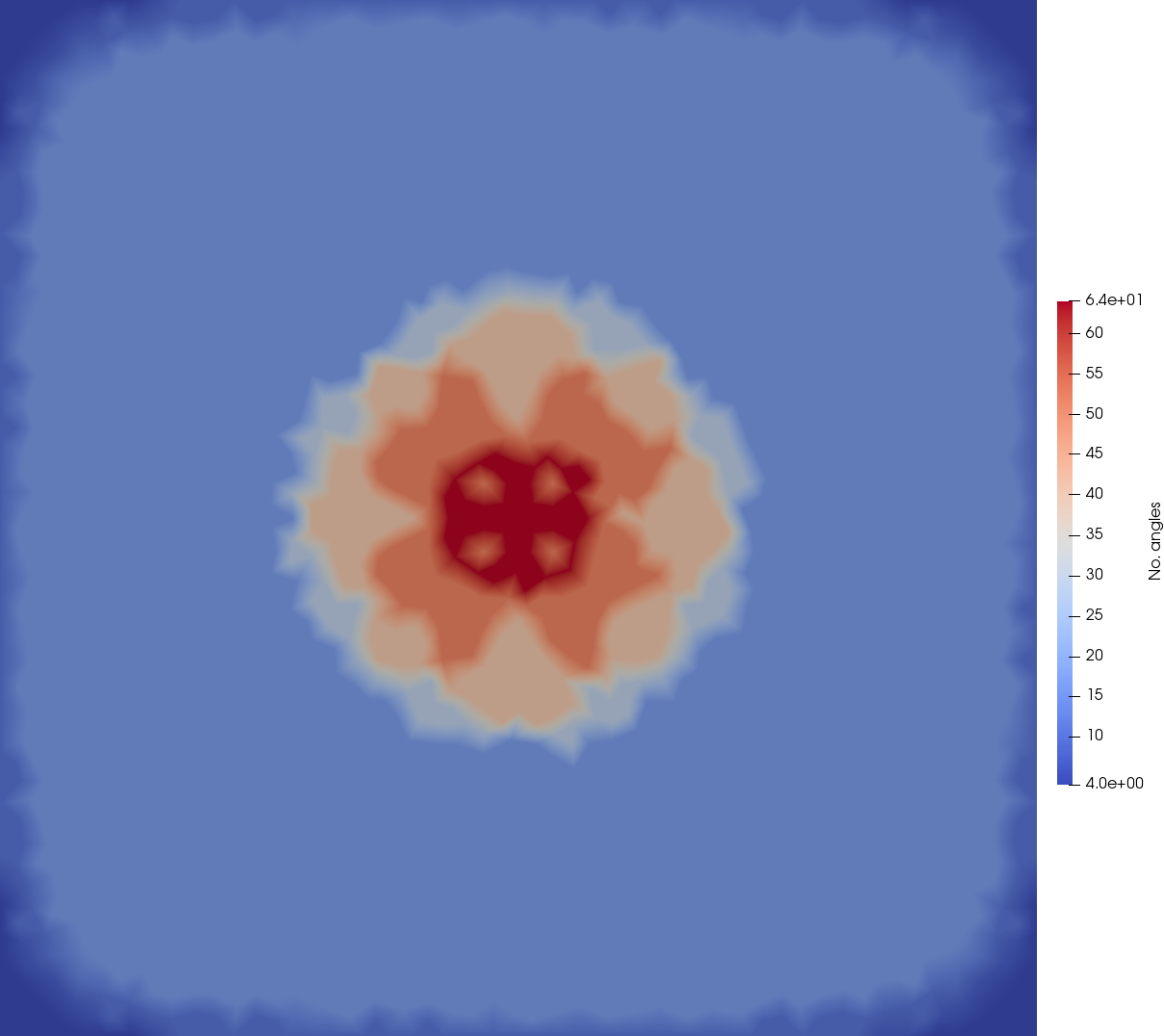}}\\
\caption{Adapt results after 3 regular adapt steps in a 2D pure scattering problem with total and scatter cross-section of 10.0 with a small source at the centre of the domain, allowing a maximum of 3 levels of regular angular refinement, with adapt steps prior to the final step solved with wavelet-based reduced tolerance to determine convergence.}
\label{fig:threshold_diffusion}
\end{figure}

We see similar results with scattering, with \tref{tab:2D_stream_adapt_diffusion_reduced} showing a substantial reduction in the number of iterations for the first two adapt steps and \fref{fig:threshold_diffusion} showing that the resulting adapted discretisation is almost identical to that in \fref{fig:no_angles_3_diffusion}. If we compare the total cost of all three adapt steps and scale by the FLOPs required to compute the matrix-free matvec for the uniform level 3 angular discretisation, the total work is reduced from 26 WUs to 17 with the reduced tolerance solves. In both cases this is less than that required to solve the uniform level 3 angular discretisation as \cite{Dargaville2023} showed this costs 41 WUs. 
\begin{table}[ht]
\centering
\begin{tabular}{ c c c | c c c c c c c}
\toprule
CG nodes & Adapt step. & NDOFs & $n_\textrm{its}$ & CC & Op Complx &  WUs$^\textrm{mf}$ & WUs$^\textrm{DG}$ & Memory\\
\midrule
2313 & 1 & 6.3\xten{4} & 2 & 4.1 & 1.7 & 4 & 7.6 & 17.2 \\
2313 & 2 & 2.4\xten{5} & 10 & 4.0 & 1.4 & 16 & 28.8 & 16.2 \\
2313 & 3 & 3.1\xten{5} & 25 & 4.1 & 1.4 & 36 & 72 & 16.3 \\
\bottomrule  
\end{tabular}
\caption{Results from using additive preconditioning on a pure scattering problem with total and scattering cross-section of 10.0 in 2D with regular angular adaptivity with a refinement tolerance of 0.001, a maximum of 3 levels of angular refinement and 3 adapt steps. The WUs listed are scaled by the nnzs in the \textit{adapted} solve at each step. The cycle and operator complexity listed are for AIRG on $\mat{M}_\Omega$ with CF splitting by Falgout-CLJP.}
\label{tab:2D_stream_adapt_diffusion_reduced}
\end{table}

\section{Conclusions}
In this work we presented an adaptive angular discretisation, with nested hierarchical P$^0$ angular elements that can be mapped to the Haar wavelet space we presented previously \cite{Dargaville2019}. Once adapted, the angular matrices required between two nodes with different angular discretisations are no longer block diagonal and hence we introduced a modified stabilisation term for use with our low-memory sub-grid scale FEM discretisation that ensured the nnzs in our adapted matrices didn't grow considerably with adaptivity. We found in both pure streaming and scattering problems that the number of nnzs grew 2-3\% above that expected from the number of adapted DOFs. This meant we could form the streaming/removal operator scalably with angular P$^0$ adaptivity and hence use AIRG and the additively preconditioned iterative method we developed \cite{Dargaville2023}. The results from this showed that we can solve our adapted problems in both streaming and scattering problems with very close to fixed work and memory. Our methods don't rely on Gauss-Seidel methods/sweeps, block-diagonal, lower triangular structure or diagonal/block scaling of our matrices and can be applied to many different discretisations. 

Given our P$^0$ discretisation is equivalent to a Haar wavelet discretisation, with adaptivity we mapped our solution to Haar space (in $\mathcal{O}(n)$) and hence tagged the angular elements that require angular refinement/de-refinement. As such we introduced the ability to robustly build up our adapted discretisation with regular adaptivity with reduced cost. We achieve this by mapping the coarse solution, $\bm{\Phi}$, to Haar space and testing the relative convergence of each wavelet coefficient at each iteration of our iterative methods. This resulted in an adapted discretisation very similar to that which would normally be produced and a large reduction in the iteration count of every adapt step prior to solving the final adapted discretisation. In a simple box test problem this reduced the cross-over point of when our adaptive process beats a uniform discretisation, down to only 5 steps with streaming or 3 with scattering. This shows the benefit of forming a P$^0$ discretisation hierarchically that can be mapped to an equivalent wavelet space, even when the solve occurs in P$^0$ space; our refinement/de-refinement is simple and we can use this to robustly reduce the cost of our adapt process with regular adaptivity. 

We also presented a CF splitting algorithm based on element agglomeration that could use the adapted angular discretisation at each spatial point to determine ``important'' directions and hence produce a semi-coarsening in the streaming limit without needing the matrix entries. We found this performed worse than typical CF splitting algorithms like Falgout-CLJP, when used to invert the streaming operator, but when used with AIRG to invert the streaming/removal operator with a large total cross-section this method scaled similarly to Falgout-CLJP, but with approximately twice the work in the solve. Given it is simple to freeze the element agglomeration after a few adapt steps, it may work out cheaper overall to use an element-agglomeration CF splitting in scattering problems with many adapt steps, but we leave examining this for future work.

Overall the combination of Falgout-CLJP CF splitting, AIRG and our additively preconditioned iterative method resulted in close to scalable work in the solve in both the streaming and scattering limit with angular adaptivity for the Boltzmann transport problem. In previous work \cite{Dargaville2019} we found that in the streaming limit the iteration count of a matrix-free multigrid method solving in Haar space tripled after only 3 adapt steps in angle; here we found the iteration count of our new method only went from 9 to 12 after 3 adapt steps. We believe this makes our iterative method an attractive choice for solving adapted discretisations of the Boltzmann transport equation. Future work will involve building an optimised version of this method in order to compare the cross-over point where our P$^0$ adaptivity results in a lower runtime than a uniform discretisation, re-using components built in prior adapt steps in order to reduce setup times and examining the performance in parallel with load-balancing.
\section*{Acknowledgments}
The authors would like to acknowledge the support of the EPSRC through the funding of the EPSRC grants EP/R029423/1 and EP/T000414/1.




\section*{References}
\bibliographystyle{model1-num-names}
\bibliography{bib_library}







\end{document}